\documentclass[aps,pra,floats,floatfix,twocolumn]{revtex4}

\usepackage{ifthen}
\usepackage{ifpdf}
\usepackage{listings}
\ifpdf
\usepackage{graphicx}
\usepackage{epstopdf}
\else
\usepackage{graphicx}
\usepackage{epsfig}
\fi

\usepackage{latexsym}
\usepackage{amsmath}
\usepackage{amssymb}
\usepackage{bm}
\usepackage{wasysym}

\newcommand{\abs}[1]{\left|#1\right|}

\newcommand{\varphiJ}{\bm{\varphi}}
\newcommand{\thetaJ}{\bm{\theta}}
\newcommand{\nJ}{\bm{n}}

\newcommand{\const}{\mbox{const}}
\newcommand{\trc}{\mbox{trace}}

\newcommand{\eexp}{\mbox{e}^}
\newcommand{\bra}{\left\langle}
\newcommand{\ket}{\right\rangle}

\newcommand{\mass}{\mathsf{m}}

\newcommand{\tbox}[1]{\mbox{\tiny #1}}

\newcommand{\amatrix}[1]{\begin{matrix} #1 \end{matrix}}

\newcommand{\be}[1]{\begin{eqnarray}\ifthenelse{#1=-1}{\nonumber}{\ifthenelse{#1=0}{}{\label{e#1}}}}
\newcommand{\beq}{\begin{eqnarray}}
\newcommand{\eeq}{\end{eqnarray}} 

\newcommand{\hide}[1]{\textcolor{red}{[hidden text]}}
\newcommand{\rmrkE}[1]{#1}

\newcommand{\Fig}[1] {Fig.~\ref{#1}} 

\begin{document}



\title{Quantum dynamics in the bosonic Josephson junction}

\author{Maya Chuchem$^1$, Katrina Smith-Mannschott$^{2,3}$, Moritz Hiller$^4$, Tsampikos Kottos$^2$, Amichay Vardi$^{5,6}$, and Doron Cohen$^1$}
\affiliation{
$^1$Department of Physics, Ben-Gurion University of the Negev, P.O.B. 653, Beer-Sheva 84105, Israel\\
$^2$Department of Physics, Wesleyan University, Middletown, Connecticut 06459, USA\\
$^3$MPI for Dynamics and Self-Organization, Bunsenstrasse10, D-37073G{\"o}ttingen, Germany\\
$^4$Physics Institute, Albert-Ludwigs-Universit{\"a}t, Hermann-Herder-Str. 3, D-79104 Freiburg, Germany\\
$^5$Department of Chemistry, Ben-Gurion University of the Negev, P.O.B. 653, Beer-Sheva 84105, Israel\\
$^6$ITAMP, Harvard-Smithsonian CFA, 60 Garden St., Cambridge, Massachusetts 02138, USA}

\begin{abstract}
We employ a semiclassical picture to study dynamics in a bosonic Josephson junction with various initial conditions.  Phase-diffusion of coherent preparations in the Josephson regime is shown to depend on the initial relative phase between the two condensates. For initially incoherent condensates, we find a universal value for the buildup of coherence in the Josephson regime. In addition, we contrast two seemingly similar on-separatrix coherent preparations, finding striking differences in their convergence to classicality as the number of particles increases. 
\end{abstract}

\maketitle

\section{Introduction}

Bose-Einstein condensates (BECs) of dilute, weakly-interacting gases offer a unique opportunity for exploring non-equilibrium many-body dynamics, far beyond small perturbations of the ground state. Highly excited states are naturally produced in BEC experiments and their dynamics can be traced with great precision and control. The most interesting possibilities lie in strong correlation effects, which imply a significant role of quantum fluctuations. 

The importance of correlations and fluctuations may be enhanced by introducing an optical lattice, that can be controlled by tuning its depth. This tight confinement decreases the kinetic energy contribution with respect to the interactions between atoms. In the tight-binding limit, such systems are described by a Bose-Hubbard Hamiltonian (BHH), characterized by the hopping frequency~$K$ between adjacent lattice sites, the on-site interaction strength~$U$, and the total atom number~$N$. The strong correlation regime is attained when the characteristic coupling parameter $u\equiv UN/K$ exceeds unity, as indicated by the quantum phase transition from a superfluid to a Mott-insulator \cite{Jaksch98,Greiner02a}.

The simplest BHH is obtained for two weakly coupled condensates (dimer).  Its dynamics is readily mapped onto a SU(2) spin problem and is closely related to the physics of superconductor Josephson junctions \cite{BJM, Leggett01}. To the lowest order approximation, it may be described by a Gross-Pitaevskii mean-field theory, accurately accounting for Josephson oscillations \cite{Javanainen86,Dalfovo96,Zapata98} and macroscopic self trapping \cite{Smerzi97}, observed experimentally in Refs.\cite{Cataliotti01,Albiez05},  as well as the equivalents of the ac and dc Josephson effect \cite{Giovanazzi00} observed in \cite{Levy07}.

Both Josephson oscillations and macroscopic self trapping rely on coherent (Gaussian) preparations, with different initial population imbalance. The mean-field premise is that such states remain Gaussian throughout their evolution so that the relative phase $\varphi$ between the two condensates remains defined. However, interactions between atoms lead to the collapse and revival of the relative phase in a process known as {\it phase diffusion} \cite{Leggett98,Wright96,Javanainen97}. 
\rmrkE{(the appropriate term is in fact {\em phase spreading}).}   
Phase diffusion has been observed with astounding precision in an optical lattice in Refs.\cite{Greiner02}, in a double-BEC system in Refs.\cite{Jo07,Schumm05,Hofferberth07}, and in a 1D spinor BEC in Ref. \cite{Widera08}. Typically, the condensates are coherently prepared, held for a varying duration (`hold time') in which phase-diffusion takes place, and are then released and allowed to interfere, thus measuring the relative coherence through the many-realization fringe visibility.  In order to establish this quantity, the experiment is repeated many times for each hold period.  

Phase-diffusion experiments focus on the initial preparation of a zero relative phase and its dispersion when no coupling is present between the condensates. However, in the presence of weak coupling during the hold time, the dynamics of phase diffusion is richer. It becomes sensitive to initial value of $\varphi$ and the loss of coherence is most rapid for $\varphi=\pi$ \cite{Vardi01,Khodorkovsky08,Boukobza09}. Here, we expand on a recent letter \cite{Boukobza09}, showing that this quantum effect can be described to excellent accuracy by means of a {\it semiclassical} phase-space picture. Furthermore, exploiting the simplicity of the dimer phase space, we derive analytical expressions
based on the classical phase-space propagation \cite{SmithMannschott09}.

Phase-space methods \cite{GardinerZoller}  have been extensively applied for the numerical simulation of quantum and thermal fluctuation effects in BECs \cite{Steel98,Sinatra01,Carusotto01,Hoffmann08,Polkovnikov03,Plimak03,Deuar06,Isella06,HKG06,Midgley09,Deuar09,Trimborn08,Mahmud}. Such methods utilize the semiclassical propagation of phase-space distributions with quantum fluctuations emulated via stochastic noise terms, 
and using a cloud of initial conditions that reflects the uncertainty of the initial quantum wave-packet. 
One particular example is the truncated Wigner approximation \cite{Sinatra01,Polkovnikov03,Isella06,Midgley09} where higher order derivatives in the equation of motion for the Wigner distribution function are neglected, thus amounting to the propagation of an ensemble using the Gross-Pitaevskii equations. 

Due to the relative simplicity of the classical phase-space of the two-site BHH, it is possible to carry out its semiclassical quantization semi-analytically \cite{Boukobza09,SmithMannschott09,Boukobza09a,Franzosi00,Graefe07} and acquire great insight on the ensuing dynamics of the corresponding Wigner distribution \cite{wignerfunc,Agarwal81}. In this work, we consider the Jopheson-regime dynamics of four different preparations,  interpreting the results in terms of the semiclassical phase-space structure and its implications on the expansion of each of these initial states in terms of the semiclassical eigenstates. We first explore the phase-sensitivity of phase-diffusion in the Josephson regime \cite{Vardi01,Boukobza09}. Then we study the buildup of coherence between two initially separated condensates \cite{Boukobza09a}, which is somewhat related to the phase-coherence oscillations observed in the sudden transition from the Mott insulator to the superfluid regime  in optical lattices \cite{Polkovnikov02,Altman02,Tuchman06}. Then we compare two coherent preparations in the separatrix region of the classical phase space, finding substantial differences in their dynamics due to their different participation numbers.

The narrative of this work is as follows \cite{nA}: The two-site Bose-Hubbard model and its classical phase-space are presented in Section~II. The semiclassical WKB quantization is carried out in Section~III. In Section~IV, we define the initial state preparations of interest and explain their phase-space representation. This is used in Section~V to evaluate their expansion in the energy basis (the local density of states), which is the key for studying the dynamics in Sections~VI to~VIII. Both the time-averaged dynamics and the fluctuations of the Bloch vector are analyzed. In particular we observe in Section~VIII that the fluctuations obeys a remarkable semiclassical scaling relation. Conclusions are given in Section X.

\section{The two-site Bose-Hubbard model}

We consider the Bose-Hubbard Hamiltonian  
for $N$ bosons in a two-site system,
\beq
\mathcal{H} \ = \ 
\sum_{i=1,2}\left[ \mathcal{E}_i \nJ_i+ \frac{U}{2} \nJ_i(\nJ_i-1)\right]  
- \frac{K}{2}(\hat{a}_2^{\dag}\hat{a}_1+ \hat{a}_1^{\dag}\hat{a}_2 ), 
\label{BHH}
\eeq
where $K$ is the hopping amplitude, $U$ is the interaction, and ${\mathcal{E}=\mathcal{E}_2-\mathcal{E}_1}$ is the bias in the on-site potentials. We use boldface fonts to mark {\em dynamical} variables that are important for the semiclassical analysis, and use regular fonts for their values. The total number of particles ${n_1+n_2=N}$ is conserved, hence the dimension of the pertinent Hilbert space 
is ${\mathcal{N} = N+1}$. Defining   
\beq
{{J}}_z \ \ &\equiv& \ \ \frac{1}{2}(\nJ_1-\nJ_2) \ \ \equiv \ \ \nJ, \\
{{J}}_+ \ \ &\equiv& \ \ \hat{a}_1^{\dag}\hat{a}_2 
\eeq
and eliminating insignificant $c$-number terms, we can re-write Eq.~(\ref{BHH}) as a spin Hamiltonian,
\be{5}
\mathcal{H} \ = \  U{{J}}_z^2 - \mathcal{E} {{J}}_z - K {{J}}_x,
\eeq
conserving the spin $J^2=j(j+1)$ with $j=N/2$. The BHH Hamiltonian thus has a {\it spherical} phase-space structure. In the absence of interaction ($U=0$) it describes simple spin precession with frequency ${\Omega=(K,0,\mathcal{E})}$. 

The quantum evolution of the spin Hamiltonian (\ref{e5})  is given by the unitary operator $\exp(-i\mathcal{H}t)$. Its classical limit is obtained by treating ${(J_x,J_y,J_z)}$ as $c$-numbers, whose  
Poisson Brackets (PB) correspond to the SU(2) Lie algebra. The classical equations of motion for any dynamical variable $A$ are derived from ${\dot{A}=-[\mathcal{H},A]_{\tbox{PB}}}$. Conservation of $J^2$ allows for one constraint on the {\it non-canonical} set of variables ${(J_x,J_y,J_z)}$.

Due to the spherical phase-space geometry, it is natural to use the non-canonical conjugate variables ${(\varphiJ,\thetaJ)}$, 
\be{91}
{{J}}_z & \equiv &  [{{J}}^2]^{1/2}\cos(\thetaJ) \\
{{J}}_x & \equiv &  [{{J}}^2]^{1/2}\sin(\thetaJ)\cos(\varphiJ)
\eeq
where $[{{J}}^2]^{1/2}$ is a constant of the motion. In the quantum mechanical Wigner picture treatment (see later sections) this constant is identified as $[(j{+}1)j]^{1/2}$, based on the derivation of section 2.3 of Ref.~\cite{wignerfunc}. For large~$j$ we use ${[{{J}}^2]^{1/2} \approx N/2}$. With these new variables the Hamiltonian takes the form
\be{9}
H(\thetaJ,\varphiJ) =   
\frac{NK}{2}\left[\frac{1}{2} u (\cos\thetaJ)^2 - \varepsilon \cos\thetaJ  - \sin\thetaJ\cos\varphiJ \right],
\eeq
where the scaled parameters are 
\be{99}
u \equiv \displaystyle{ \frac{NU}{K}},  
\hspace*{2cm}
\varepsilon \equiv \displaystyle{\frac{\mathcal{E}}{K}}.
\eeq
The structure of the underlying classical phase space is determined 
by the dimensionless parameters $u$ and $\varepsilon$. For strong interactions ${u>1}$ 
the phase space includes a figure-eight shaped separatrix (\Fig{f1}), provided  $\abs{\varepsilon}<\varepsilon_c$, where \cite{WN00,SmithMannschott09}
\be{11}
\varepsilon_c=\left(u^{\frac{2}{3}}-1\right)^{\frac{3}{2}}.
\eeq
This separatrix splits phase space into three integrable regions:  a `sea' of Rabi-like trajectories and two interaction-dominated nonlinear `islands'.  For zero bias the separatrix is a symmetric $8$ shaped figure 
that encloses the two islands, and the relevant energies are \cite{ELS85}
\be{12}
E_{-}  =&  -(1/2)NK   &=\mbox{ground energy},
\\
E_{\tbox{x}}  =&  +(1/2)NK     &=\mbox{separatrix},
\\ \label{e14}
E_{+}  =&  (1/4)[u{+}(1/u)]NK   &=\mbox{top energy}.
\eeq
Looking at the phase-space structure as a function of $u$, the two islands emerge once ${u>1}$.   
For ${u>2}$, they encompass the North and the South poles, yielding macroscopic self-trapping \cite{Smerzi97}, 
and for very large ${u\gg1}$ they cover most of the Northern and the Southern hemispheres respectively. 
Note that for $u\gg1$, the expression ${E_{+}\approx (N/2)^2U}$, reflects the cost of localizing all the particles 
in one site, compared to equally-populated sites.

As an alternative to Eq.~(\ref{e9}), it is possible to employ the relative number-phase representation, using the canonically conjugate variables ${(\varphiJ,\nJ)}$, with the Hamiltonian,
\be{1414}
\mathcal{H} \ \ = \ \ 
U  \nJ^2 
\ - \ \mathcal{E}  \nJ   
\ - \ K\sqrt{(N/2)^2-\nJ^2} \ \cos(\varphiJ) .
\eeq
In the ${|n| \ll (N/2)}$ region of phase space, one obtains 
the Josephson Hamiltonian, which is essentially the Hamiltonian of a pendulum 
\beq
\mathcal{H}_{\tbox{Josephson}} \ \ = \ \ E_C(\nJ-n_{\varepsilon})^2 \ - \ E_J\cos(\varphiJ)
\eeq
with $E_C=U$ and $E_J=KN/2$ while $n_{\varepsilon}$ is linearly related to $\mathcal{E}$.
The Josephson Hamiltonian ignores the spherical geometry, 
and regards phase space as {\em cylindrical}.  
Accordingly it captures much of the physics if the motion is restricted 
to the  small-imbalance slice of phase-space, 
as in the case of equal initial populations and strong interactions. 
However, for other regimes it is an over-simplification 
because it does not correctly capture the global topology.

\section{The WKB quantization}

We begin by stipulating the procedure for the semiclassical quantization of the spin spherical phase-space. In the ${(\varphi,n)}$ representation, the area element is ${d\Omega=d\varphi dn}$,  so that the total phase-space area is~$2\pi N$ with Planck cell ${h=2\pi}$. 
Alternatively, using the ${(\varphi,\theta)}$ coordinates, the area element is ${d\Omega=d\varphi d\cos\theta}$, so that the total area is~$4\pi$. Consequently   
\beq
h \ = \ \mbox{Planck cell area in steradians} \  = \ \frac{4\pi}{\mathcal{N}} .
\eeq
Within the framework of the semiclassical picture,  eigenstates are associated with {\em stripes} of area $h$ that are stretched along contour lines ${\mathcal{H}(\varphi,\theta)=E}$ of the classical Hamiltonian $\mathcal{H}$. The associated WKB quantization condition is 
\beq
A(E_{\nu}) = \left(\frac{1}{2} + \nu\right)h, 
\ \ \ \ \ \ \nu=0,1,2,...
\eeq
where $A(E)$ is defined as the phase-space area 
enclosed by an $E$ contour in steradians:
\beq
A(E) \ \ \equiv \ \  \iint \Theta(E-\mathcal{H}(\varphi,\theta))  \ d\Omega,
\eeq
while the area  of phase space in Planck units is $A(E)/h$.
The frequency of oscillations at energy $E$ is 
\be{18}
\omega(E) \ \ \equiv \ \ \frac{dE}{d\nu} \ \ = \ \ \left[\frac{1}{h}A'(E)\right]^{-1}.
\eeq

Consider the semiclassical quantization of the Hamiltonian (\ref{e9}). 
We distinguish three regimes of interaction strength.  
Assuming ${\varepsilon=0}$, these are 
\beq
\mbox{Rabi regime:} &  u<1 & \mbox{(no islands)}
\\
\mbox{Josephson regime:} & 1<u<N^2 & \mbox{(see Fig.~\ref{f1})}
\\
\mbox{Fock regime:} & u>N^2 & \mbox{(empty sea)}
\eeq
In the Fock regime, the area of the sea becomes 
less than a single Planck cell, and therefore cannot support any eigenstates.
Our interest throughout this paper is mainly in the Josephson regime where neither 
the $K$ term nor the $U$ term can be regarded as a small perturbation in the Hamiltonian. 
This regime, characteristic of current atom-interferometry experiments, 
is where semiclassical methods become most valuable.

In the WKB framework the spacing between energies equals the characteristic 
classical frequency at this energy. If the interaction is zero ($u=0$), 
the energy levels are equally spaced and there is only one frequency, namely the Rabi frequency 
\beq
\omega_K \ \ = \ \ K
\eeq
For small interaction  (${0<u<1}$) the frequencies around the two stable fixed points  are slightly modified to the plasma frequencies: 
\beq
\omega_{\pm} \ \ = \ \  \omega(E_{\pm}) 
\ \ = \ \ \sqrt{(K \pm NU)K}.  
\eeq
If the interaction is strong enough ($u\gg1$), 
the oscillation frequency near the 
minimum point can be approximated as 
\beq
\omega_J \ \ = \ \  \omega(E_-) 
\ \ \approx \ \  \sqrt{NUK} 
\ \ = \ \  \sqrt{u} \ \omega_K,
\eeq
while at the top of the islands we have:
\beq
\omega_{+}  \ \ = \ \  \omega(E_{+})  \ \ \approx \ \  NU  \ \ = \ \ u \ \omega_K.
\eeq
The associated approximations for the phase-space area 
in the three energy regions (bottom of the sea, 
separatrix, and top of the islands) are respectively  
\beq
\frac{1}{h}A(E)  &=&  \left(\frac{E-E_{-}}{\omega_J}\right),  
\\ \label{eq29}
\frac{1}{h}A(E) &=&  \frac{1}{h}A(E_{\tbox{x}}) 
+ \frac{1}{\pi}\left(\frac{E-E_{\tbox{x}}}{\omega_J}\right) \log\left|\frac{NK}{E-E_x}\right|,
\\
\frac{1}{h}A(E) &=& \frac{4\pi}{h} - \left(\frac{E_{+}-E}{U}\right)^{1/2},  
\eeq
In the last expression the total area
of the two islands ${(E_{+}{-}E)/U}$ 
should be divided by two if one wants to obtain 
the area of a {\em single} island.   
A few words are in order regarding the derivation of Eq.~(\ref{eq29}).
In the vicinity of the unstable fixed point the contour lines 
of the Hamiltonian are ${Un^2-(NK/4)\varphi^2=E{-}E_{\tbox{x}}}$, 
where for convenience the origin is shifted ($\varphi{=}\pi\mapsto0$).
Defining $a^2=4(E{-}E_{\tbox{x}})/NK$  
the area of the region above the separatrix is $2(NK/4U)^{1/2}[\mathcal{A}(a)-\mathcal{A}(0)]$
where $\mathcal{A}(a)$ is the integral over $\sqrt{\varphi^2+a^2}$. 
The result of the integration is ${\mathcal{A}(a)-\mathcal{A}(0)=a^2\log(b/a)}$, 
with some ambiguity with regard to $b\sim1$ which is 
determined by the outer limits of the integral where the hyperbolic 
approximation is no longer valid.

Away from the separatrix, the WKB quantization recipe 
implies that the local level spacing at energy $E$
is~$\omega(E)$ given by Eq.~(\ref{e18}).  
In particular, the low energy levels have spacing $\omega_J$, 
while the high energy levels are doubly-degenerate with spacing $\omega_{+}$.
In the vicinity of the separatrix we get 
\be{30}
\omega(E) \ \ \approx \ \ \left[\frac{1}{\pi}\log\left|\frac{NK}{E-E_x}\right|\right]^{-1} \omega_J.
\eeq
Using the WKB quantization condition, we find that the level spacing 
at the vicinity of the separatrix ($E\sim E_{\tbox{x}}$)   
is {\em finite} and given by the expression   
\be{31}
\omega_{\tbox{x}} \ \ = \ \ \left[\log\left(N/\sqrt{u}\right)\right]^{-1} \omega_J.
\eeq
Using an iterative procedure one finds that at 
the same level of approximation 
the near-separatrix energy levels are   
\beq
E_{\nu} = E_{\tbox{x}} +  \left[\frac{1}{\pi}\log\left|\frac{N/\sqrt{u}}{\nu{-}\nu_{\tbox{x}}}\right|\right]^{-1}  (\nu{-}\nu_{\tbox{x}}) \ \omega_J,
\eeq
where $\nu_{\tbox{x}}=A(E_{\tbox{x}})/h$. \Fig{f1} demonstrates 
the accuracy of the WKB quantization, and of the above approximations.

\section{The initial preparation and its phase-space representation - the Wigner function}

Our approach for investigating the dynamics of various initial preparations relies on the Wigner-function formalism for spin variables, developed in  Refs.~\cite{Agarwal81,wignerfunc}. Each initial preparation is described as a Wigner distribution function over the spherical phase space. The dynamics is deduced from expanding the initial state in terms of the semiclassical eigenstates described in Sec~III. In this section we specify the Wigner distribution for the preparations under study whereas the following section presents the eigenstate expansion of each of these four initial wavepackets, evaluated semiclassically.

To recap the phase-space approach to spin \cite{Agarwal81,wignerfunc}, the Hilbert space of the BHH has the dimension $\mathcal{N}=2j{+}1$, and the associated space of operators has the dimensionality $\mathcal{N}^2$. According to the Stratonovich-Wigner-Weyl correspondence (SWWC) \cite{Stratonovich56}, any observable $A$ in this space, as well as the probability matrix  of a spin$^{(j)}$ entity, can be represented by a real sphere$^{(2j)}$ function $A_{\tbox{W}}(\Omega)$.
The sphere$^{(2j)}$ is spanned by the $Y^{\ell m}(\Omega)$ functions with ${\ell \le 2j}$, and the practical details regarding this formalism can be found in Refs.~\cite{Agarwal81,wignerfunc}.  
The SWWC allows to do exact quantum calculation in a classical-like manner. A few examples for Wigner functions pertinent to this work, are displayed in \Fig{f2}. Expectation values are calculated  as in classical statistical mechanics: 
\be{1234}
\text{trace}[\hat{\rho} \ \hat{A} ] \ \ = \ \ 
\int\frac{d\Omega}{h}{\rho_{\tbox{W}}(\Omega) A_{\tbox{W}}(\Omega) }
\eeq
In particular the Wigner-Weyl representation of the identity operator is~$1$, and that 
of~$J_x$ is as expected $[(j{+}1)j]^{1/2}\sin(\theta)\cos(\varphi)$ \cite{wignerfunc}.
We adopt the convention that~$\rho_{\tbox{W}}$ is normalized with respect to the 
measure~$d\Omega/h$,   allowing to handle on equal footing a cylindrical phase space 
upon the re-identification  ${d\Omega=d\varphi dn}$ and ${h=2\pi}$.

Within this phase-space representation, the Fock states $|n\rangle$ are represented 
by stripes along constant~$\theta$ contours (see e.g. left panel of \Fig{f2}). 
The ${|n{=}N\rangle}$ state (all particles in one site) is a Gaussian-like wave packet 
concentrated around the NorthPole. From this state, we can obtain a family of  
spin coherent states (SCS) ${|\theta,\varphi\rangle}$ via rotation.


In what follows, we explore the dynamics of the following experimentally-accessible preparations (see \Fig{f2}), the first being a Fock state, whereas the last three are spin coherent states: 
\begin{itemize}

\item 
{\em TwinFock preparation:}
The $n{=}0$ Fock preparation. 
Exactly half of the particles are in each side of the double well.  
The Wigner function is concentrated along the equator $\theta=\pi/2$.


\item 
{\em Zero preparation:} 
Coherent $(\theta{=}\pi/2, \varphi{=}0)$ preparation, located entirely in the (linear) sea region. 
Both sites are equally populated with definite $0$ relative phase. 
The minimal wave-packet is centered at $(n=0,\varphi=0)$.

\item 
{\em Pi preparation:}
Coherent $(\theta{=}\pi/2, \varphi{=}\pi)$ on-separatrix preparation. 
The sites are equally populated with $\pi$ relative phase.  
The minimal wave-packet is centered at $(n=0,\varphi=\pi)$.

\item
{\em Edge preparation:}
Coherent $\varphi\ne\pi$ on-separatrix preparation.
The minimal wave-packet is centered on the 
separatrix but away from the saddle point on the $\varphi{=}0$ side. 

\end{itemize}

The Wigner function of an SCS resembles a {\em minimal}  
Gaussian wave-packet, and it should satisfy 
\beq
\int\rho_{\tbox{W}}(\theta,\varphi) \frac{d\Omega}{h} 
= 
\int [\rho_{\tbox{W}}(\theta,\varphi)]^2 \frac{d\Omega}{h} = 1.
\eeq
This requirement helps to determine the phase space 
spread without the need to use the lengthy algebra of Refs.~\cite{Agarwal81,wignerfunc}. 
For the Fock coherent state ${|n{=}N\rangle}$,   
that is centered at the NorthPole ($\theta=0$), 
one obtains:
\beq
\rho_{\tbox{W}}^{(\psi)}(\theta,\varphi) 
\approx  
2\eexp{-\frac{\mathcal{N}}{2}\theta^2} ~.
\eeq
For the coherent states centered around the Equator, 
it is more convenient to use ${(\varphi,n)}$ coordinates, 
e.g. the $\varphi=0$ coherent state is well approximated as, 
\be{4141}
\rho_{\tbox{W}}^{(\psi)}(n,\varphi) 
\ \ \approx \ \ 
\frac{1}{ab} 
\eexp{ -\frac{\varphi^2}{2 a^2} - \frac{n^2}{2b^2}} ~,
\eeq
with $a=1/\sqrt{2\mathcal{N}}$ and $b=\sqrt{\mathcal{N}/2}$.
Shifted versions of these expressions describe the ${\varphi=\pi}$ and the ``Edge" preparations. 
The Wigner function of a Fock state ${\psi=|\mathsf{n}\rangle}$  is semiclassically approximated as
\be{4242}
\rho_{\tbox{W}}^{(\psi)}(n,\varphi) 
\ \ \approx \ \  
\delta(n-\mathsf{n}),
\eeq
whereas the Wigner function
of an eigenstate is semiclassically approximated 
by a micro-canonical distribution: 
\be{40}
\rho_{\tbox{W}}^{(\nu)}(n,\varphi) 
\ \ \approx \ \  
\omega(E_{\nu}) \ \delta(\mathcal{H}(\varphi,n)-E_{\nu}). 
\eeq
In general none of the above listed initial states is an eigenstate of the BHH, 
but rather a superposition of BHH eigenstates. Consequently their Wigner function deforms 
over time (see e.g. \Fig{f3}) and the expectation values of observables 
become time dependent (see e.g.\Fig{f4}), as discussed in later sections.

\section{The initial preparation and its eigenstate expansion - local density of states}

\begin{figure}[h!]

\centering
\includegraphics[width=0.4\hsize]{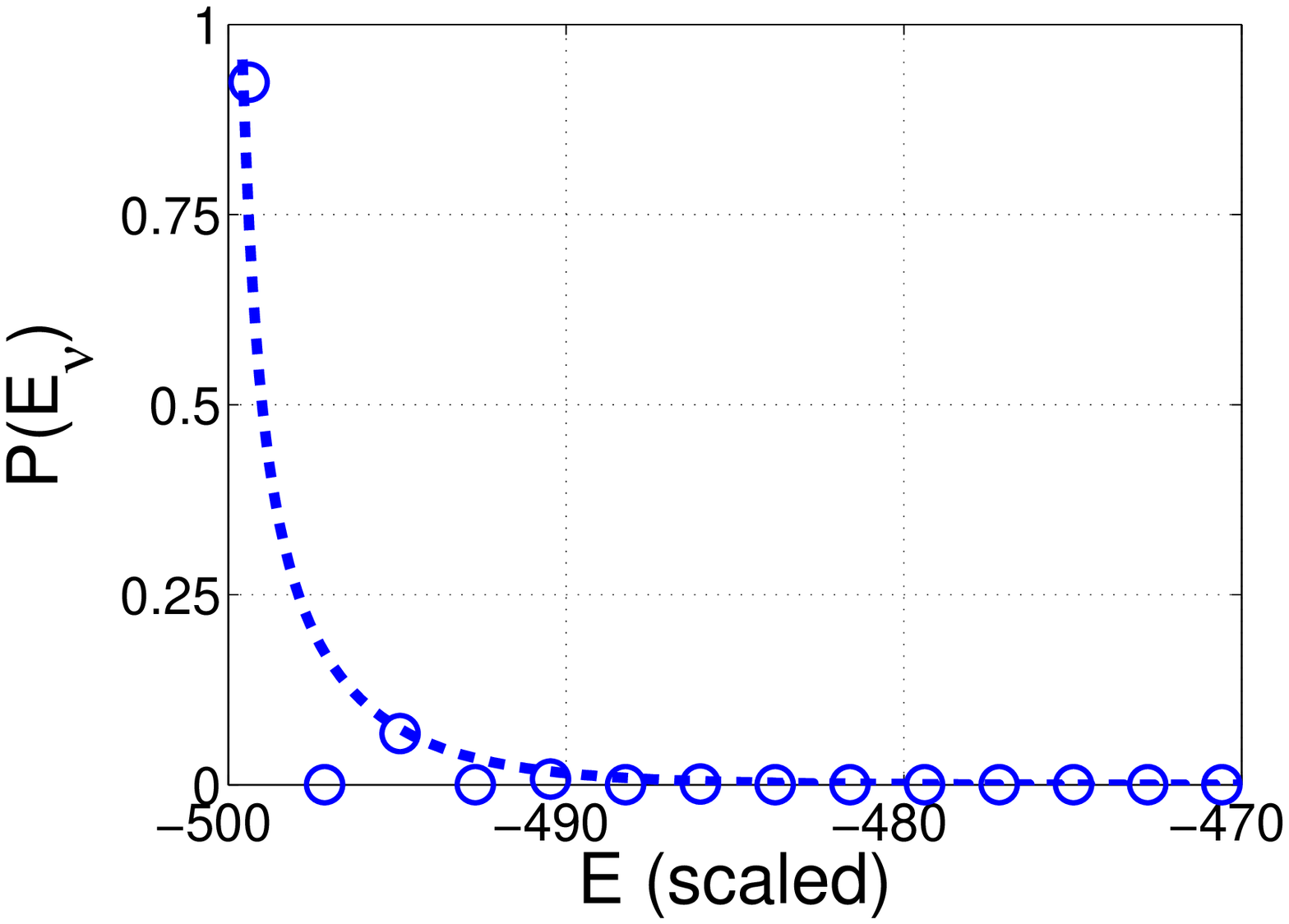} 
\includegraphics[width=0.4\hsize]{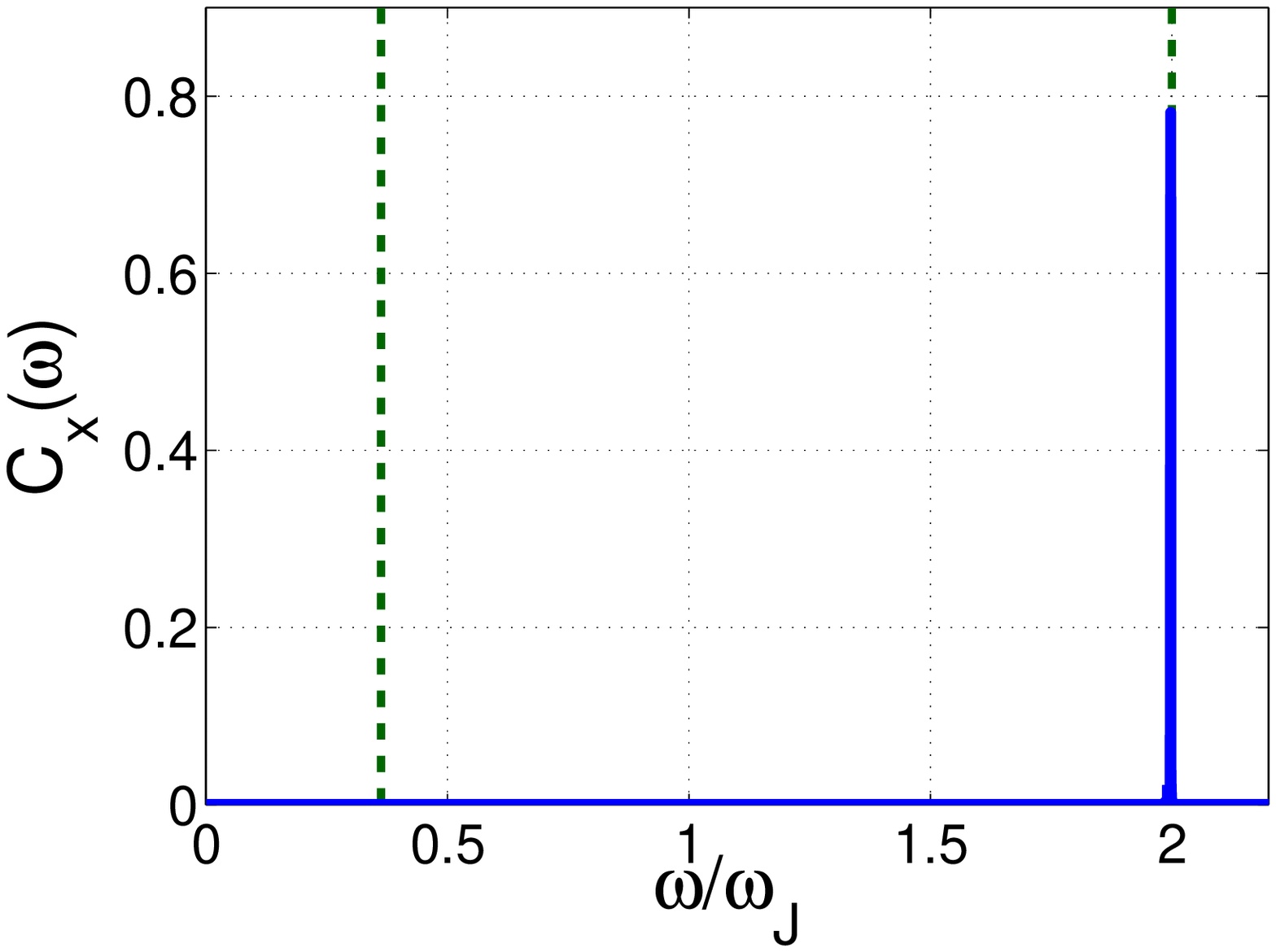}\\
\includegraphics[width=0.4\hsize]{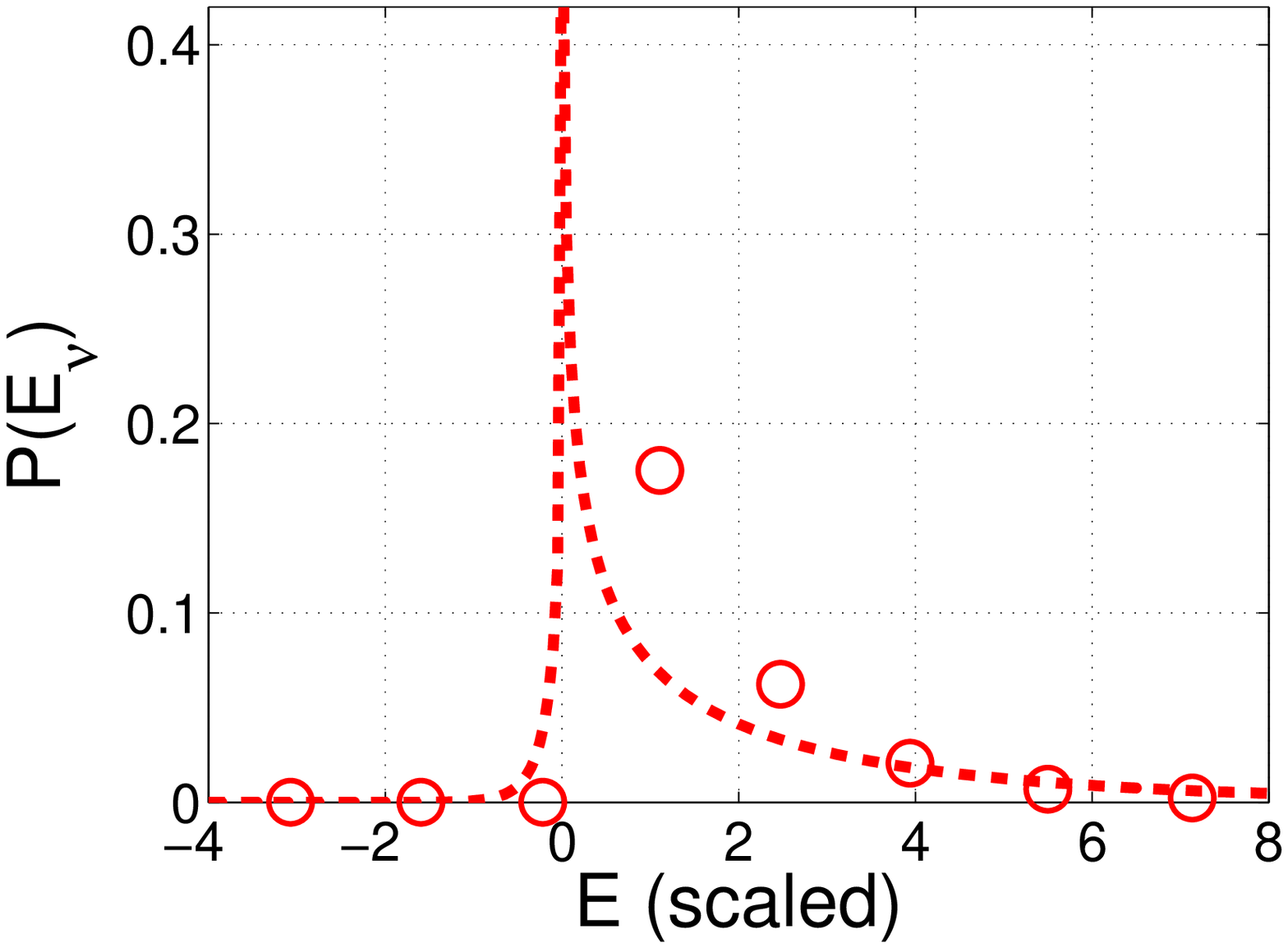} 
\includegraphics[width=0.4\hsize]{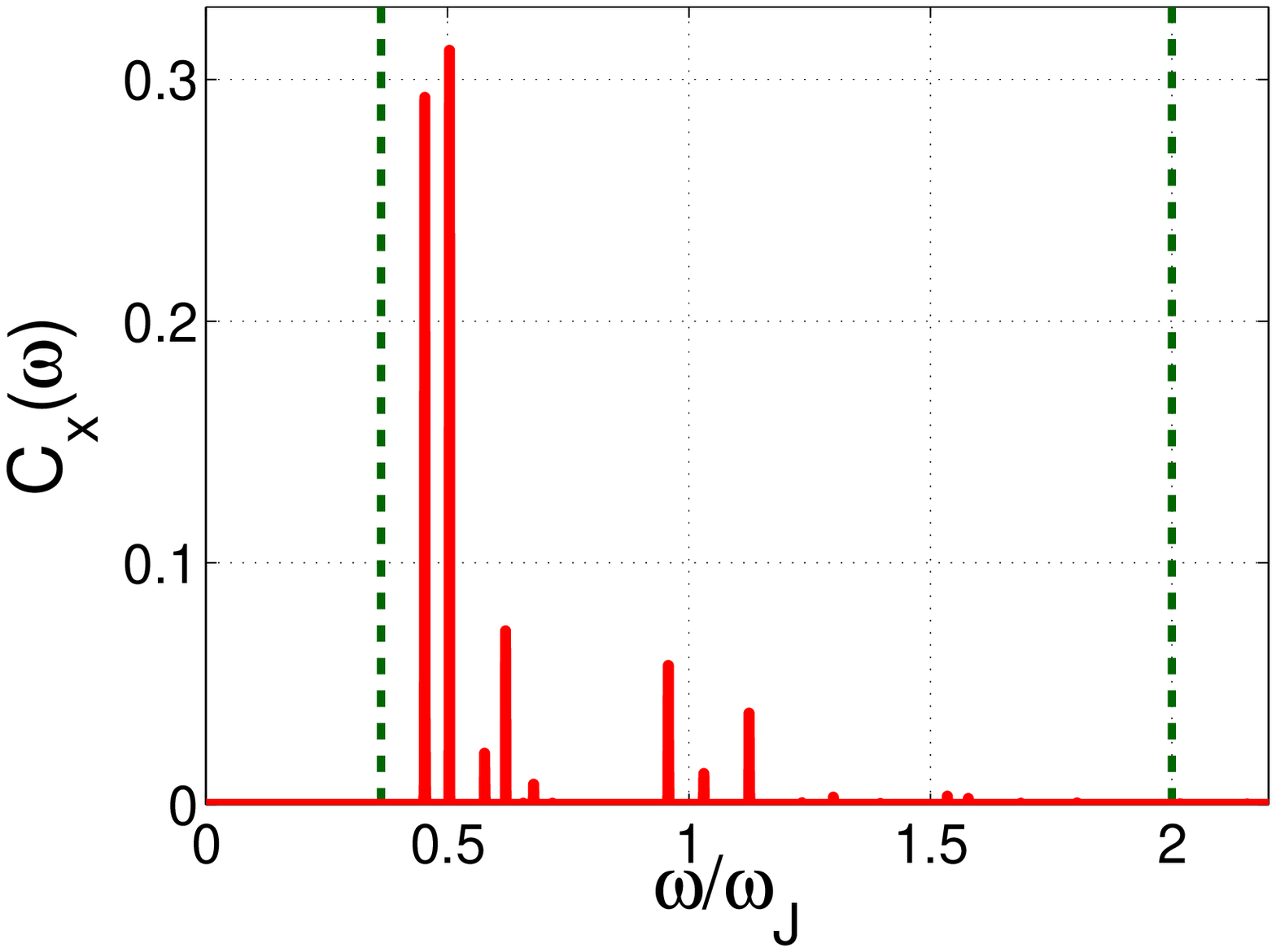}\\ 
\includegraphics[width=0.4\hsize]{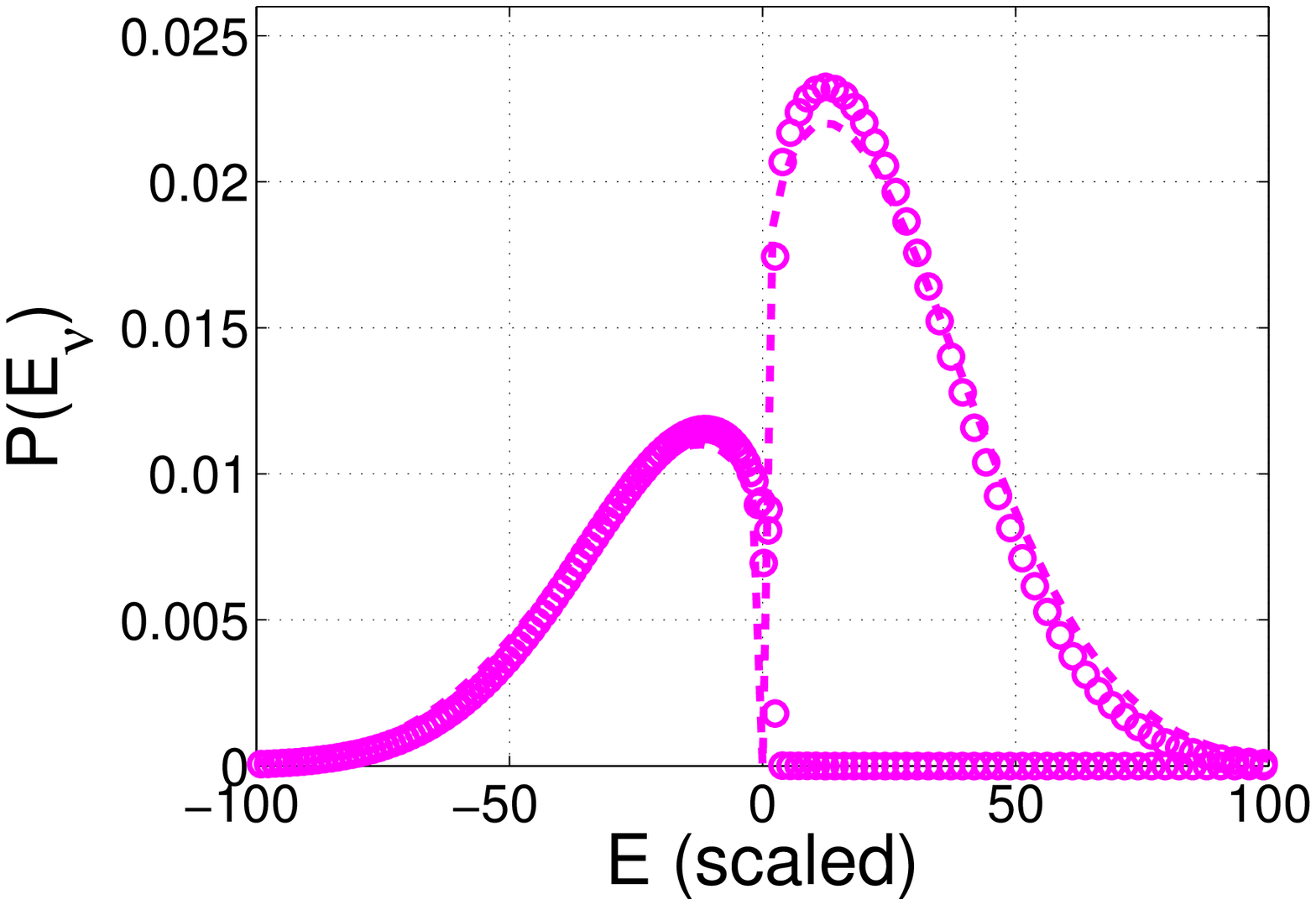} 
\includegraphics[width=0.4\hsize]{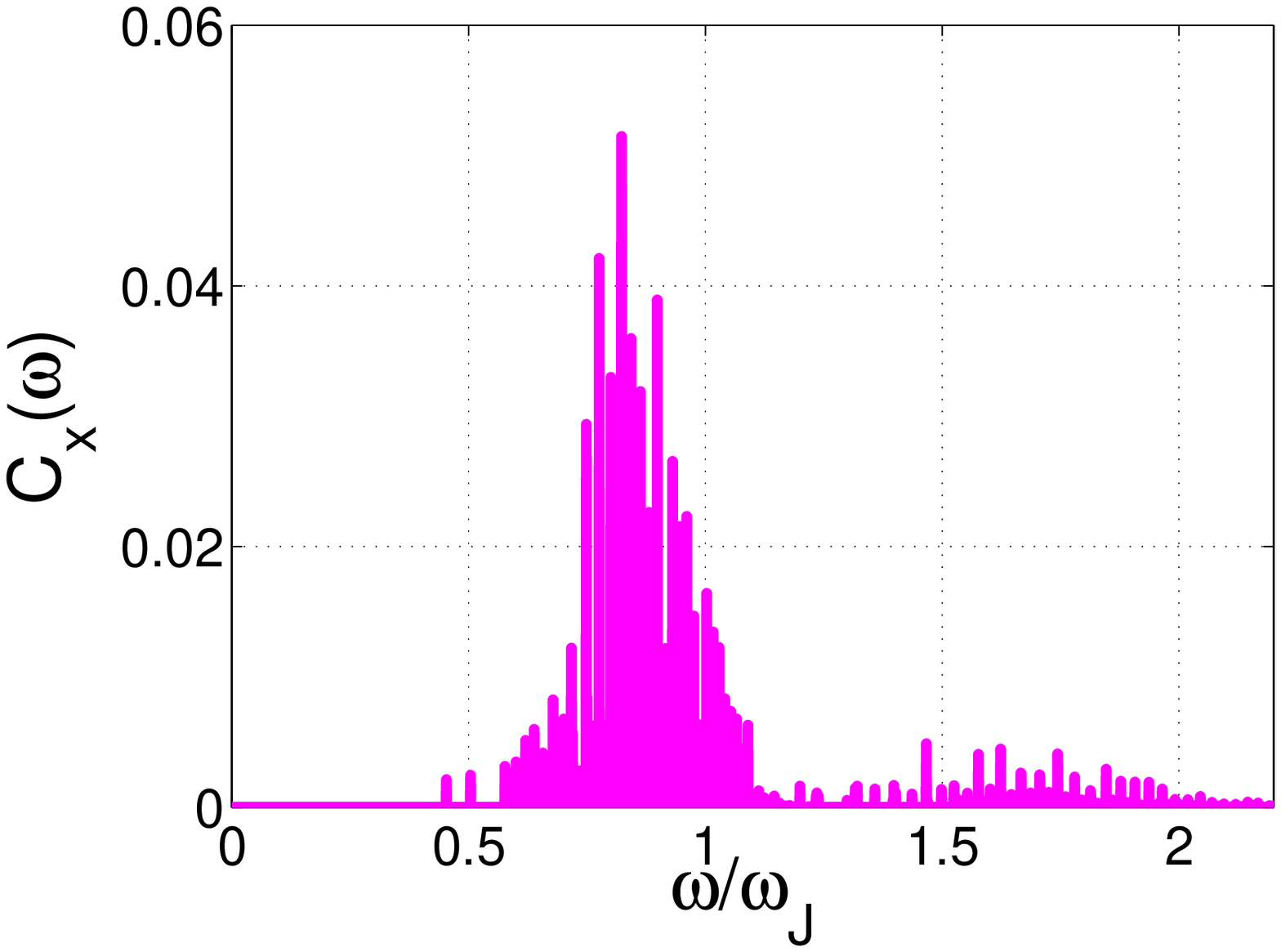}\\
\includegraphics[width=0.4\hsize]{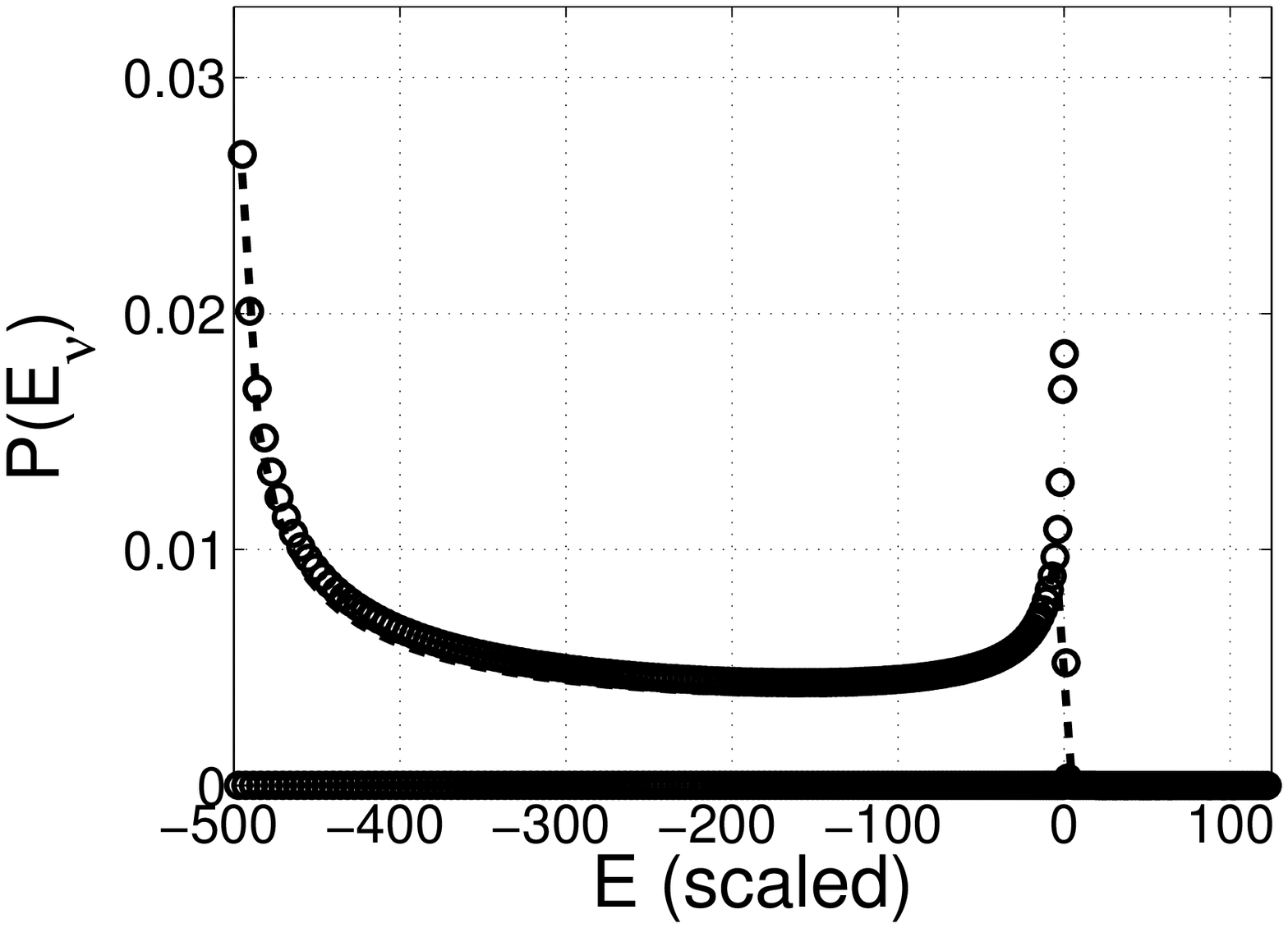} 
\includegraphics[width=0.4\hsize]{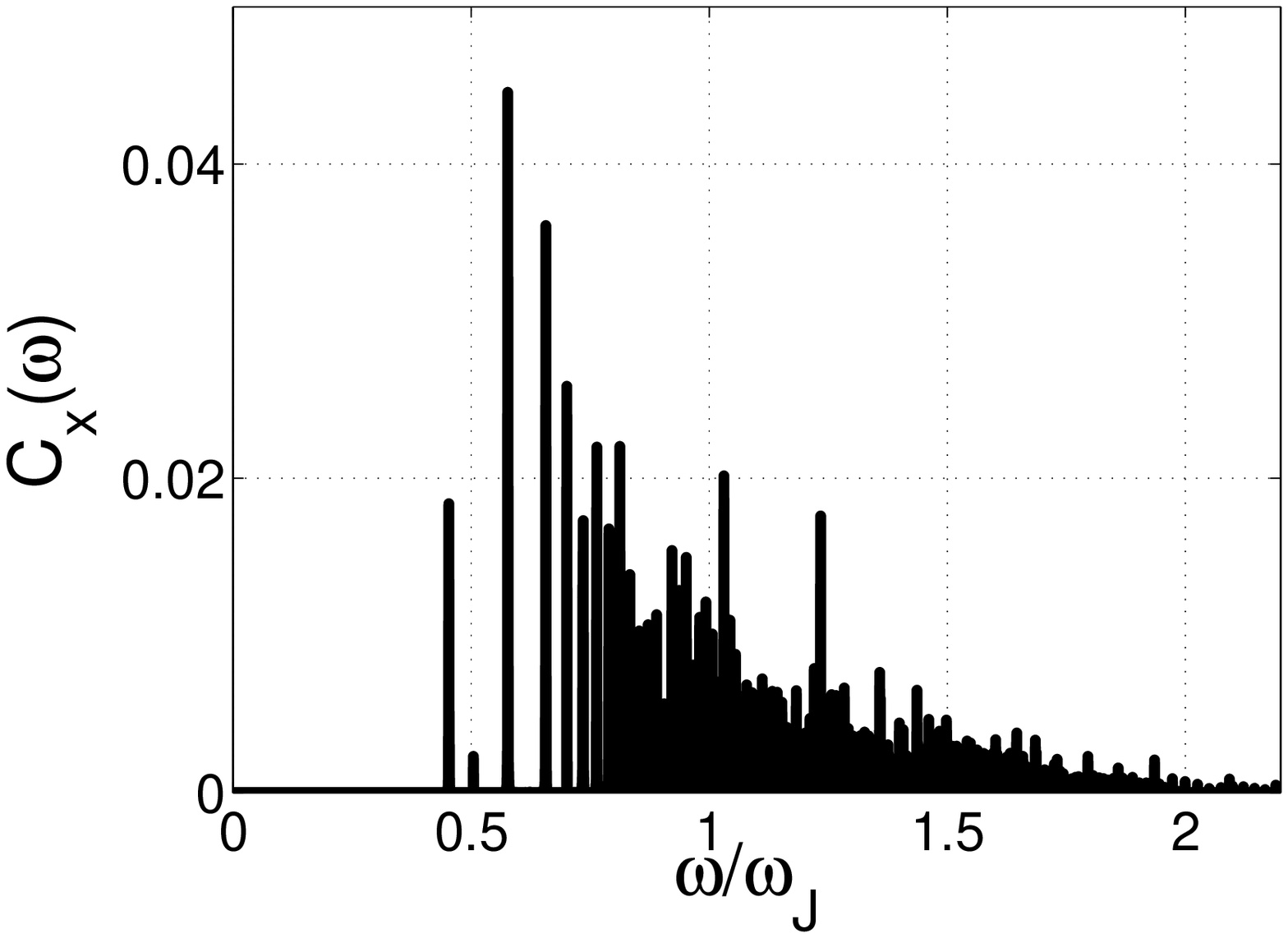}

\caption{(Color online)
The LDOS (left) and the spectral content of the fluctuations (right) 
of $N=500$  bosons with $u=4$, for Zero, Pi, Edge, and TwinFock preparations (top to bottom).
The horizontal axes are  $E-E_{\tbox{x}}$ and $\omega/\omega_J$.
The lines in the LDOS figures are based on a semiclassical analysis (see text), 
while the circles are from the exact quantum calculation. 
Note that due to the mirror symmetry of the Zero preparation  
the expected frequency should approach $2\omega_J$, 
while for the Pi preparation it is bounded from below by $2\omega_{\tbox{x}}$ 
(both frequencies are indicted by vertical dashed lines).
Note also the outstanding difference between the spectral support 
of Zero and Pi preparations compared with continuous-like support 
in the case of Edge and Fock preparations. 
}
\label{f5}
\end{figure}

Having set the stage by defining the phase-space representation of eigenstates in Sect. III and of the initial conditions in Sect. IV, the evolution of any initial preparation is uniquely determined by its eigenstate expansion. Thus, in order to analyze the dynamics, we now evaluate  the {\it local density of states} (LDOS)  with respect to the preparation $\psi$ of the system:  
\be{-1}
\mbox{P}(E_{\nu}) 
&=&
|\langle E_{\nu} | \psi \rangle|^2 = 
\trc(\rho^{(\nu)}\rho^{(\psi)})
\\ \label{e42}
&=&
\int \rho_{\tbox{W}}^{(\nu)}(\Omega) \ \rho_{\tbox{W}}^{(\psi)}(\Omega) \frac{d\Omega}{h}.
\eeq
If $\psi$ is mirror symmetric the 
above expression should be multiplied either 
by~0 or by~2 in the case of odd/even eigenstates. In \Fig{f5}, we plot the LDOS associated with Pi, Edge, Zero and TwinFock preparations.
The applicability of semiclassical methods to calculate the LDOS has been numerically demonstrated for the case of a three site (trimer) model in \cite{HKG06}.  By contrast, the simpler two site (dimer) model under consideration, offers an opportunity for deriving exact expressions by substituting Eqs.~(\ref{e4141})-(\ref{e40}) into Eq.~(\ref{e42}) and evaluating the integrals under the appropriate approximations.  For clarity, we summarize below the main results of this analytic evaluation, with the details  given in App.~\ref{sD}. 
 
Consider first the Pi and Edge preparations. The Wigner distributions of both lie on the separatrix and are hence concentrated around the energy $E_{\tbox{x}}$. Yet, their line shapes are strikingly different: For an Edge preparation we have,
\beq
\mbox{P}(E) 
\ \propto \ 
\frac{\omega(E)}{\omega_J} \ \exp\left[-\frac{1}{\mathcal{N}}\left(\frac{E-E_{\tbox{x}}}{\omega_J}\right)^2 \right]~ ,
\eeq
featuring a {\it dip} at $E \sim E_{\tbox{x}}$. The calculation in the Pi case leads to 
\beq
\mbox{P}(E) 
\ \propto \ 
\frac{\omega(E)}{\omega_J} \ \mbox{Bessel}\left[\frac{E-E_{\tbox{x}}}{NU}\right] 
\eeq
(the exact expression can be found in App.~\ref{sD}). This latter line shape features a {\it peak} at $E \sim E_{\tbox{x}}$, because the Bessel function compensates the logarithmic  suppression by~$\omega(E)$.  Consequently, as seen below, the fluctuations associated with on-separatrix motion differ dramatically, depending on where the SCS wave-packet is launched.

Analytic results are obtained in App.~\ref{sD}, also for the Zero and TwinFock preparations.  In the latter case the result is:
\be{4848}
\mbox{P}(E) 
\ \propto \ 
\frac{\omega(E)}{NK}
\left[1-\left(\frac{2E}{NK}\right)^2\right]^{-1/2}~.
\eeq
As shown in  \Fig{f5}, the above semiclassical expressions agree well with the exact numerical results.  

The LDOS  determines the spectral content of the dynamics. Consider first the characteristic oscillation frequency of dynamical variables. Away from the separatrix, it is given by the classical estimate $\omega_{\tbox{osc}} \approx \omega(E)$. For example, for a Zero preparation, we have ${\omega_{\tbox{osc}}  = \omega_J}$. While the classical frequency vanishes on the separtrix, we still obtain a finite result for near-separatrix preparations (Pi,Edge) because the wave-packet has a finite width, and because $\omega_{\tbox{x}}$ provides a lower bound on $\omega_{\tbox{osc}}$. In any case the result becomes $h$ dependent.  To be specific, consider the Pi and the Edge preparations separately. In both cases the width of the wave-packet is $\Delta n=\sqrt{N/2}$. In case of the Pi preparation it occupies an energy range ${\Delta E = U\Delta n^2 \propto N}$, while in the Edge preparation case ${\Delta E = v_n \Delta n \propto N^{1/2}}$,   
where $v_n \approx \omega_J$ is identified  as the velocity of the phase space points in this region. Using Eq.~(\ref{e30}) we find 
\be{46}
\omega_{\tbox{osc}}  \approx  \left\{ \amatrix{1 \cr 2} \right\} \times \left[\log\left(\frac{N}{u}\right)\right]^{-1} \omega_J,
\eeq
where the additional factor of ``2" applies to the Edge preparation:
This factor is due to the different dependence of $\Delta E$ on $N$ 
in the two respective cases. On top we might have an additional 
factor of~$2$ due to mirror symmetry (see discussion of the dynamics 
at the end of the next section).
Note that as $u/N$ exceeds unity, the distinction between Pi, Zero, and Edge preparations blurs
because the wave-packets becomes wider than the width of the separatrix region, until at the Fock regime, all three consist of the same island levels. Thus in this limit, we get for $\omega_{\tbox{osc}}$ essentially the same result as in the (negligable $K$) Fock regime:  the width of the wave-packet is $\Delta n=\sqrt{N/2}$, and the dispersion relation $\omega(E)=2Un$ gives  the standard separated-condensates phase-diffusion frequency \cite{Greiner02,Jo07},    
\be{47}
\omega_{\tbox{osc}}  \ \approx  \ \left(\frac{u}{N}\right)^{1/2} \omega_J \ = \ U\sqrt{N}.
\eeq
It should be clear that, both, in Eq.~(\ref{e46}) and in Eq.~(\ref{e47})
$u/N$ should be accompanied with a numerical prefactor that should 
be adjusted because the notion of ``width" is somewhat ill defined 
and in general depends on the precise details of the numerical procedure, 
which we discuss in the next section.

\begin{figure}[h!]
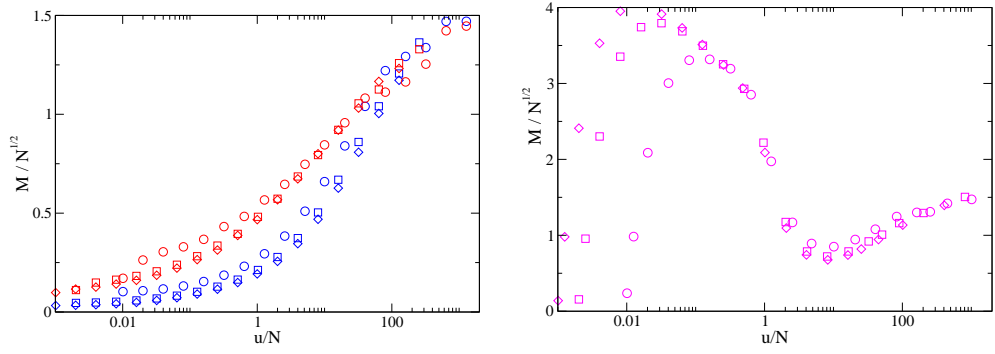


\centering
\includegraphics[clip,width=0.35\textwidth]{M_phi_pi} 
\ \ \ \ 
\includegraphics[clip,width=0.35\textwidth]{M_edge} 

\caption{(Color online) 
The participation number $M$ as determined from the LDOS  
for $N=100 \,(\circ),500\, (\Box),$ and $1000\,(\diamond)$ particles.
The left panel contains the Zero (lower set in blue) 
and Pi (upper set in red) preparations, 
while the Edge preparation is presented in the right panel.
Note the different vertical scale. 
In the crudest approximation we expect in the Edge case ${M \sim N^{1/2}}$, 
while in the Pi case  ${M \ll N^{1/2}}$ as long as ${(u/N)\ll 1}$  
}
\label{f6}
\end{figure}

For the analysis of the temporal fluctuations  
it is crucial to determine the number of eigenstates 
that participate in the wave-packet superposition. 
This is given by the LDOS {\it participation number},
\be{48}
M \ \ = \ \ \left[\sum_{\nu} \mbox{P}(E_{\nu})^2\right]^{-1}.
\eeq  
See \Fig{f6} for numerical results. The participation number
can be roughly estimated as ${M=\Delta E/\omega_{\tbox{osc}}}$, 
where $\Delta E$ is the energy width of the wave-packet
and $\omega_{\tbox{osc}}$ is the mean level spacing.  
In the case of a Pi preparation 
\be{49}
M \approx \left[\log\left(\frac{N}{u}\right)\right] \sqrt{u},
\eeq
while for an Edge preparation we find
\be{50}
M \approx \left[\log\left(\frac{N}{u}\right)\right] \sqrt{N}.
\eeq
Note that $M/\sqrt{N}$ is a function of the semiclassical 
ratio~$(N/u)^{1/2}$ between the energy width of the wave-packet 
and the width of the separatrix region.
The expected scaling is confirmed 
by the numerical results of \Fig{f6}.  
The above approximations for $M$ assume~${u/N<1}$ and 
are useful for the purpose of rough estimates.   

In the next sections we analyze the temporal 
fluctuations of some observables. The associated Fourier 
power spectrum (\Fig{f5}, right panels) is related to 
the LDOS content of the wave-packet superposition. 
Both the  characteristic frequency (\Fig{f7}) 
and the spectral {\em spread} of the frequencies
can be estimated from the $\omega_{\tbox{osc}}$ and the $M$
that are implied by the above LDOS analysis.

\section{Dynamics (i) - the time evolution of the Bloch vector}

After describing the semiclassical phase-space picture and using it to determine the spectral structure of the initial preparations, we now turn to the ensuing dynamics. At present, most experiments on the bosonic Josephson system, measure predominantly quantities related to the  one-body reduced probability matrix, defined via the expectation values ${S_i=(2/N) \langle J_i\rangle}$ as,   
\beq
\rho^{[1]}_{ji} 
\ \ = \ \  \frac{1}{N}\langle \hat{a}_i^{\dagger} \hat{a}_j \rangle   
\ \ = \ \  \frac{1}{2}(\hat{\bm{1}} + \bm{S} \cdot \hat{\bm{\sigma}})_{ji} ,
\eeq
where ${\bm{S} = (S_x,S_y,S_z)}$ is the Bloch vector and $\hat{\bm{\sigma}}$ is composed of  Pauli matrices.  The population imbalance and relative phase between the two sites, determined by direct imaging and the position of interference fringes \cite{Albiez05}, are given respectively by, 
\beq
\mbox{OccupationDiff} &=& N \, S_z,
\\
\mbox{RelativePhase} &=&  \arctan(S_x, S_y),
\eeq
whereas single-particle purity is reflected by the measures,
\beq
\mbox{OneBodyPurity} &=&  (1/2) \left[ 1 {+} S_x^2 {+} S_y^2 {+} S_z^2 \right],
\\
\mbox{FringeVisibility} &=&  \left[ S_x^2 + S_y^2 \right]^{1/2}.
\eeq
In particular the FringeVisibilty, given by the transverse component of the Bloch vector, corresponds to the visibility of fringes, averaged over many realizations \cite{Schumm05,Hofferberth07}. Loosely speaking it reflects the phase-uncertainty of the state. Coherent states have maximum OneBodyPurity. Of these, equal-population coherent states have maximal FringeVisibility with the smallest phase-variance.   
Starting from a coherent preparation, single-particle purity can be diminished over time due to nonlinear effects (see below) or due to interaction with environmental degrees of freedom (decoherence).  By contrast, Fock states carry no phase information, but interactions may lead to their dynamical phase-locking over time and to the buildup of FringeVisibility (see below).

\begin{figure}[h!]

\centering
(a) \hspace{0.45\hsize} (b) \hspace{0.45\hsize} \\

\includegraphics[width=0.45\hsize]{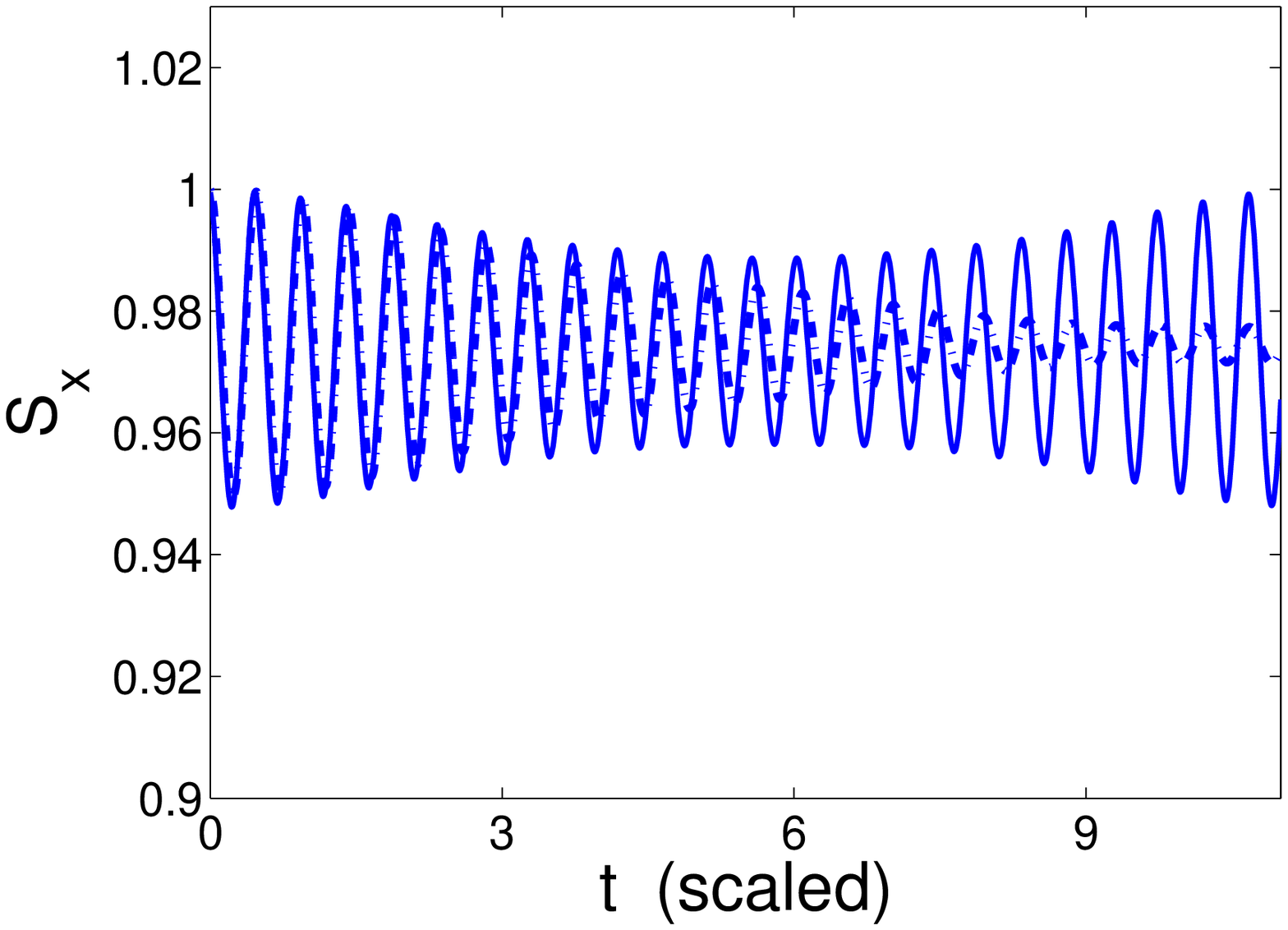} 
\includegraphics[width=0.45\hsize]{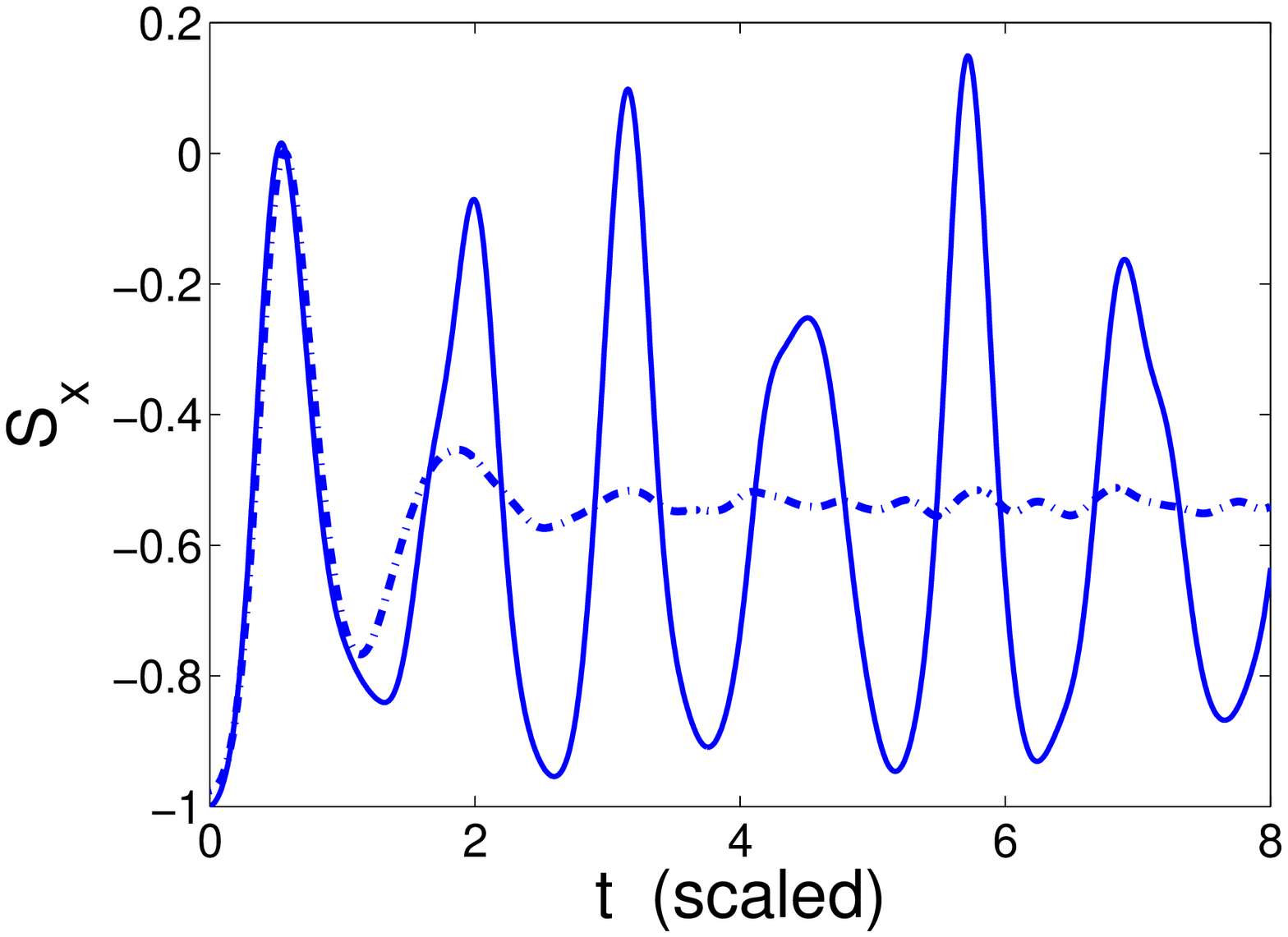} \\

(c) \hspace{0.45\hsize} (d) \hspace{0.45\hsize} \\

\includegraphics[width=0.45\hsize]{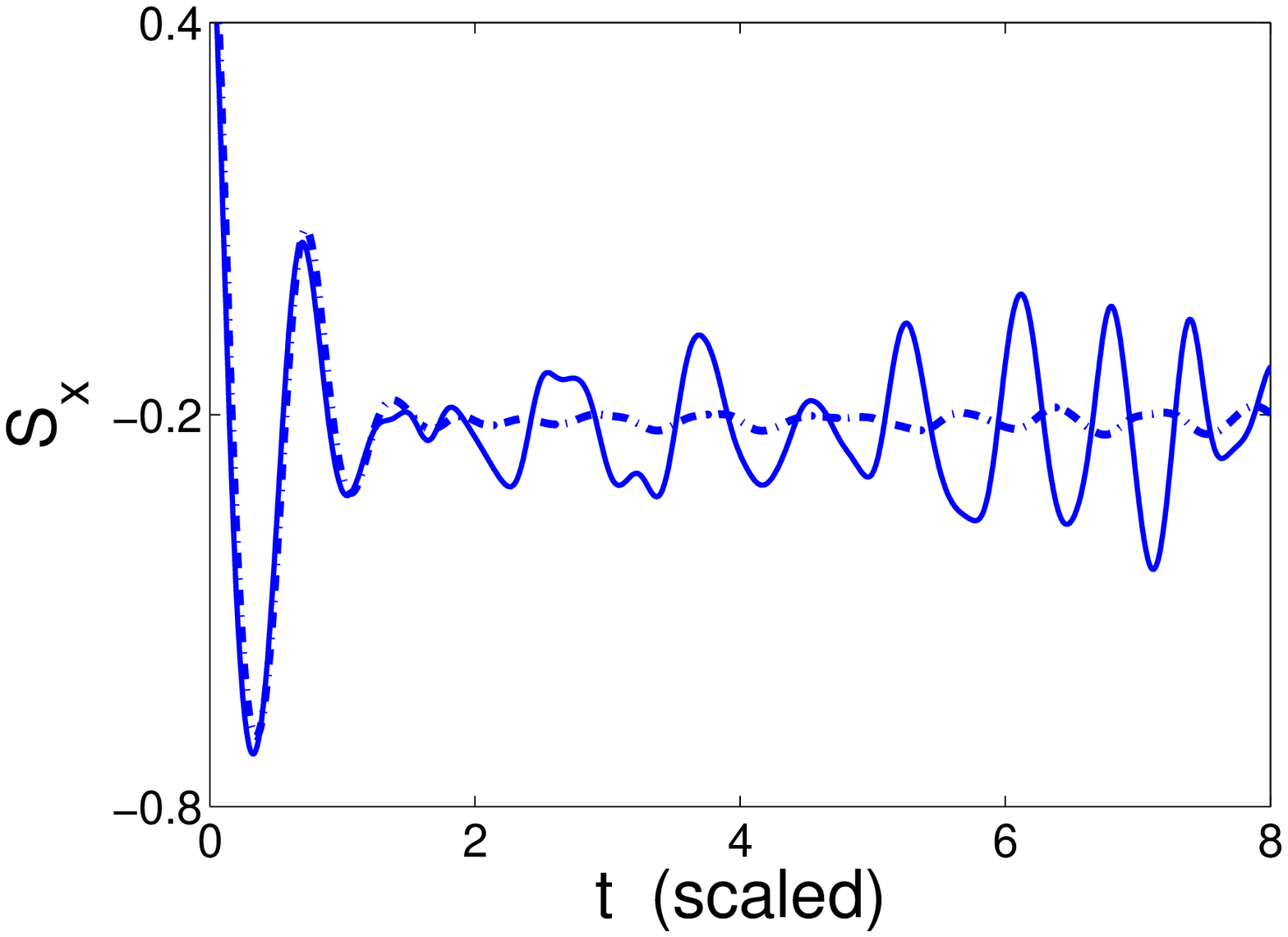} 
\includegraphics[width=0.45\hsize]{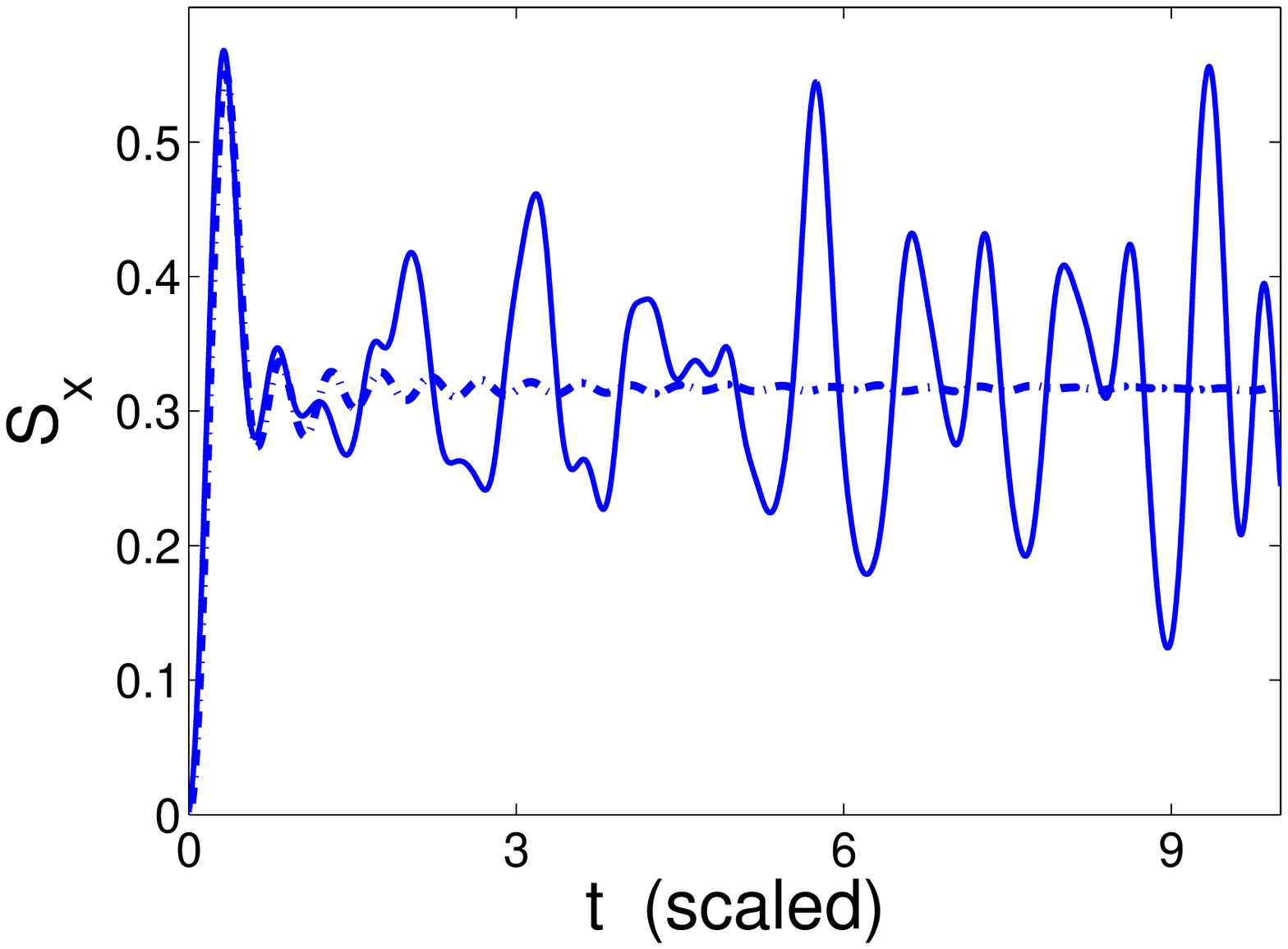}

\caption{(Color online) 
The variation of $S_x(t)$ with time for $N{=}40$ particles with $u=5$,   
for Zero (a), Pi (b), Edge (c), and TwinFock (d) preparations. 
Note the different vertical scale. 
The dashed-dotted lines are based on semiclassical simulation.
Note that the fluctuations of a semiclassical preparation 
always die after a transient, which should be contrasted 
with both the classical (single trajectory) and 
quantum (superposition of ${M>1}$ eigenstates) behavior.      
}
\label{f4}
\end{figure}

We carry out numerically-exact quantum simulations where the state $|\psi\rangle$ is propagated 
according to $|\psi(t)\rangle=\exp(-it\mathcal{H})|\psi\rangle$, where the Hamiltonian 
is given by Eq.(\ref{e5}) which is equivalent to Eq.(\ref{BHH}).  
The evolved state after time~$t$ can be visualized using its Wigner function. 
For example, the time evolution of the initial TwinFock state is illustrated in \Fig{f3}, 
with the Wigner function of an evolved state shown in \Fig{f3}(a). 
Comparison is made in \Fig{f3}(b) to the Liouville propagation for the same duration, 
of the corresponding cloud of points, according to the classical equations,
\be{2233}
\dot{S}_x &=& u S_{z} S_{y} ~, \\
\dot{S}_y &=& -(1+uS_{x}) S_{z} ~,\\
\dot{S}_z &=& -S_{y} ~,
\eeq
where time has been rescaled (${t:= Kt}$). Good quantum-to-classical correspondence 
is observed for short-time simulation (e.g. see \Fig{f4}). 
In \Fig{f3}(c) we plot the resulting occupation statistics, 
which can be regarded as a projection of the phase space distribution 
(classical, dash-dotted line) or Wigner function (quantum, solid line), 
namely $\mbox{P}_t(n)= |\langle n|\psi(t) \rangle|^2 =\trc(\rho^{(n)}\rho^{(\psi(t))})$.
Our main interest is in the FringeVisibility, and hence in 
\be{5555}
S_x(t) = \frac{2}{N} \langle {{J}}_x \rangle 
= \frac{[(j{+}1)j]^{1/2}}{j} \langle \sin(\thetaJ)\cos(\varphiJ) \rangle ~. 
\eeq
The prefactor in the last equality is implied by Eq.~(\ref{e91}), 
and cannot be neglected if the number of particles is small. 
In \Fig{f4} we plot $S_x(t)$ for the four preparations defined in Sec.~IV, 
comparing semiclassical results (dash-dotted lines) 
to the numerical full quantum calculation (solid lines). 
For all equatorial preparations (Zero,Pi,TwinFock), $S_x$ is in fact the FringeVisibility, 
because $S_y(t)=0$ identically throughout the evolution. 
As a general observation, the semiclassical simulation captures well 
the short-time transient evolution and the long time average $\overline{S_x}$. 
For example, for the initial TwinFock preparation (\Fig{f4}d), 
it reproduces the universal Josephson-regime FringeVisibility of~$\sim 1/3$, 
resulting from the dynamical smearing of the Wigner distribution function 
throughout the linear sea region of phase-space \cite{Boukobza09}.

In what follows we would like to determine the long time average $\overline{S_x}$ and the 
power spectrum $\tilde{C}(\omega)={\rm FT}[f(t)]$ of the  temporal fluctuations $f(t)=S_x(t)-\overline{S_x}$.
(FT denotes Fourier Transform).  Characteristic power-spectra for the pertinent preparations are shown in the right panels of \Fig{f5}. We characterize the fluctuations by their typical frequency $\omega_{\tbox{osc}}$, by their spectral support (discrete or continuous-like), and by their RMS value:  
\be{59}
\mbox{RMS}[S_x] = [\overline{f(t)^2}]^{1/2} = \left[\int \tilde{C}(\omega)  d\omega\right]^{1/2},
\eeq
The dependence of $\omega_{\tbox{osc}}$, and $\overline{S_x}$, and $\mbox{RMS}[S_x]$
on the dimensionless parameters $(u,N)$ is illustrated in Figs.~\ref{f7}-\ref{f8}.
It should be realized that the observed frequency of the $S_x(t)$ 
oscillations is in fact $2\omega_{\tbox{osc}}$ due to the mirror symmetry of the observable.

\section{Dynamics (ii) - fringe visibility in the Josephson regime}
 
In the Fock regime, the FringeVisibility of an initial coherent $\varphi$ preparation decays to zero:
the initial Gaussian-like distribution located at ${(\theta{=}\pi/2,\varphi)}$ is stretched along the equator, 
leading to increased relative-phase uncertainty with fixed population-imbalance. This {\em phase spreading} process 
is known in the literature as {\em `phase diffusion'} 
\cite{Leggett98,Wright96,Javanainen97,Greiner02,Widera08}.
By contrast, a TwinFock preparation is nearly an eigenstate of the BHH in this regime, 
and its zero FringeVisibility remains vanishingly small from the beginning.

In the {\em Josephson regime} the dynamics of the single-particle coherence is more intricate. 
In this section we  discuss the evolution of $\bm{S}(t)$ within the framework of the semiclassical approximation. 
As a first step, we disregard fluctuations and recurrences and address only the time-averaged dynamics. 
In particular, we determine the long-time average of the FringeVisibility, 
which for the TwinFock, Pi, and Zero preparations is given by $\overline{S_x(t)}$, as noted after Eq.~(\ref{e5555}).

\begin{figure}[h!]
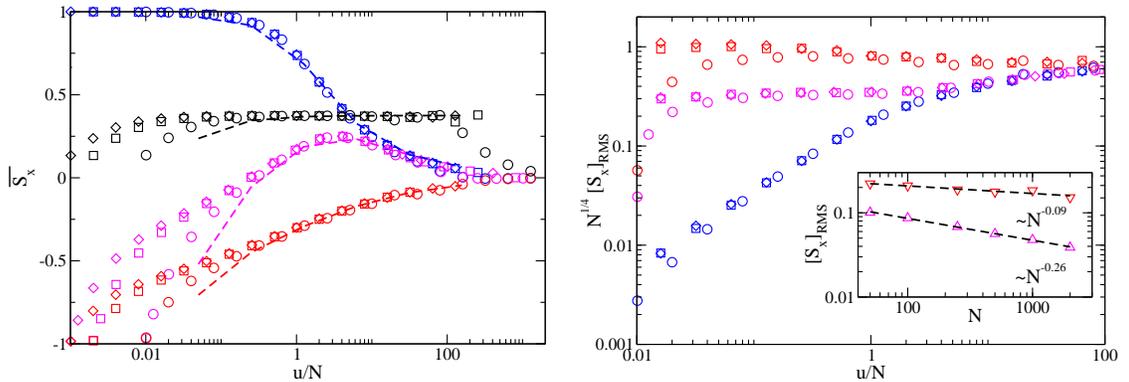


\centering
\includegraphics[clip,width=0.4\textwidth]{s_x}
\ \ \ 
\includegraphics[clip,width=0.4\textwidth]{RMS_s_x_scaled} 

\caption{(Color online)
{\em Left:} The long-time average of $S_x(t)$ versus $u/N$
for $N=100 \,(\circ),500\, (\Box),$ and $1000\,(\diamond)$ particles.
The preparations are (upper to lower sets of data points):  
Zero (blue), TwinFock (black), Edge (magenta), and Pi (red). 
The symbols are used for the quantum results and
the dashed lines are the semiclassical prediction for fourty particles.
Note that the scaling holds only in the Josephson regime ${1\ll u \ll N^2}$,
and therefore, for a given $u/N$ range, becomes better for large~$N$.
{\em Right:} 
The long time RMS of $S_x(t)$ for the three coherent preparations 
(lower to upper sets): Zero (blue), Edge (magenta), and Pi (red). 
The implied $N^{1/4}$ scaling based on Eq.~(\ref{e73})) is confirmed.  
In the inset, the RMS of $S_x(t)$ for Edge ($\triangle$) and Pi ($\triangledown$) preparations
is plotted versus~$N$ while~${u=4}$ is fixed. The dashed lines are power-law fits 
that nicely agree with the predictions of Eq.~(\ref{e74}).
}    

\label{f8}
\end{figure}

Numerical results for the long-time average and for the RMS of the fluctuations as a function of~$u/N$ 
are presented in \Fig{f8} and further discussed below.  The main observations regrading the dynamics 
in the Josephson regime are: 
\begin{itemize}
\item
The TwinFock preparation of fully-separated condensates develops phase-locking at $\varphi{\sim}0$, 
with FringeVisibility ${\overline{S_x(t)} \approx 1/3}$ \cite{Boukobza09a}.
\item  
Starting from a SCS preparation, phase-diffusion becomes phase sensitive. 
The Zero preparation is phase-locked, while the coherence of the Pi preparation 
is partially lost \cite{Vardi01,Boukobza09,Boukobza10}, exhibiting huge fluctuations.
\item
The Edge preparation exhibits distinct behavior, 
that neither resembles the Zero nor the Pi preparations, 
involving sign reversal of $\overline{S_x(t)}$. 
\end{itemize}

In the remaining part of this section we quantify these observations by finding the long-time average $\overline{S_x(t)}$ based on simple phase-space considerations. In the Josephson regime, the value of  $\overline{S_x(t)}$  for a coherent preparation should be determined by the ratio between its ${\Delta n \approx \sqrt{N/2}}$ width and the  width of the separatrix region ${\Delta n \approx \sqrt{NK/U}}$, i.e.
\be{63}
\mbox{the semiclassical ratio} \ \ = \ \ (u/N)^{1/2} ~.
\eeq
This ratio determines the long-time {\em phase-space distribution} of the evolving semiclassical cloud:  
In the case of a TwinFock preparation this cloud fills the entire sea region; 
in the "Zero" case it is confined to an ellipse within the sea region; 
and in the "Pi" case it stretches along the separatrix and therefore resembles a micro-canonical distribution. 
The projected phase-distribution $\mbox{P}(\varphi)$ is determined accordingly. 
Disregarding a global normalization factor we get  
\be{6464}
\mbox{P}(\varphi) \approx& \exp[-\varphi^2/(4u/N)] 
&\mbox{[Zero]} ~,
\\ \label{eq62}
\mbox{P}(\varphi) \approx& [(u/N)+\cos^2(\varphi/2)]^{-1/2}  
&\mbox{[Pi]} ~,
\\ \label{eq63}
\mbox{P}(\varphi) \approx& \cos(\varphi/2)  
&\mbox{[TwinFock]} ~.
\eeq
A few words are in order regarding the derivation of the above expressions.
The Zero case preparation is represented by the Gaussian 
of Eq.(\ref{e4141}) whose major axis is ${\Delta n \approx (N/2)^{1/2}}$. 
This Gaussian evolves along the contour lines of 
the Hamiltonian ${\mathcal{H}(n,\varphi)=Un^2+(NK/2)\cos(\varphi)}$.
After sufficiently long time the evolving distribution 
still has the same ${\Delta n}$, but because of the spreading 
its other major axis, as determined by the equation ${U\Delta n^2=(NK/4)\Delta \varphi^2}$, 
becomes ${\Delta \varphi \approx (2U/K)^{1/2}}$ leading to Eq.(\ref{e6464}). 
The TwinFock preparation is represented by Eq.(\ref{e4242}).
After sufficiently long time the evolving distribution 
fills the whole sea ${\mathcal{H}(\theta,\varphi)<E_{\tbox{x}}}$.
The equation that describes this filled sea  
can be written as ${n<n_{\tbox{x}}(\varphi)}$, where 
\be{644}
n_{\tbox{x}}(\varphi) = \sqrt{\frac{NK}{2U}\left(1+\cos(\varphi)\right)} ~.
\eeq
The projection of area under ${n<n_{\tbox{x}}(\varphi)}$  
is simply ${\mbox{P}(\varphi) \propto n_{\tbox{x}}(\varphi)}$, leading to Eq.(\ref{eq63}).
The Pi case preparation is represented by the Gaussian that is 
located on the separatrix. After sufficiently long time the evolving 
distribution is stretched along ${n \sim n_{\tbox{x}}(\varphi)}$, 
and looks like ${\delta(\mathcal{H}(\theta,\varphi)-E_{\tbox{x}})}$. 
If we neglected the finite width of this distribution we 
would obtain ${\mbox{P}(\varphi) \propto 1/n_{\tbox{x}}(\varphi)}$, 
which is divergent at ${\varphi \sim \pi}$. But if we take  
into account the ${\Delta n \approx \sqrt{N/2}}$ width 
of the preparation, which is effectively like adding $(N/2)$ under the 
square root of Eq.(\ref{e644}), then we get Eq.(\ref{eq62}).

As implied by the Wigner-Weyl picture one can get 
from $\mbox{P}(\varphi)$ the long time average $\overline{S_x}$
through the integral 
\beq
\overline{S_x} \approx \int \cos(\varphi) \, \mbox{P}(\varphi) d\varphi ~.
\eeq
The calculation is straightforward leading to  
\beq
\overline{S_x} \approx
& \exp[-(u/N)] \ \ \ 
&\mbox{[Zero]},
\\ 
\overline{S_x} \approx
& -1-4/\log\left[\frac{1}{32}(u/N)\right] \ \ \ 
&\mbox{[Pi]},
\\
\overline{S_x} \approx
& 1/3 \ \ \ 
&\mbox{[TwinFock]}.
\eeq
These expressions agree with the numerics of \Fig{f8}, and confirm the predicted scaling with $u/N$.

\section{Dynamics (iii) - long time fluctuations}

To complete our phase-space characterization of the Bloch vector dynamics, we need to address the long-time fluctuations $f(t)$ from the average value $\overline{S_x}$. As demonstrated in \Fig{f4}, the evaluation of $\overline{S_x}$ could be done to excellent accuracy based on the semiclassical propagation of phase-space distributions according to purely {\it classical} equations of motion. This corresponds to the truncated Wigner approach of quantum optics \cite{Sinatra01,Polkovnikov03,Isella06,Midgley09}, retaining only the leading Liouville term in the equation of motion for the Wigner distribution. Obviously one can not guarantee that the remaining Moyal-bracket terms, which are initially ${\cal O}(1/N)$, will remain small throughout the evolution. This is the source of the fluctuations observed in \Fig{f4}. While their characterization goes beyond the lowest-order truncated Wigner semiclassics, we still can estimate them based on the phase-space LDOS expansion, as described below.

The two-site BHH has essentially {\em one}~degree~of~freedom since both the energy and the number of the bosons is conserved. Therefore, away from the separatrix, level spacing is determined by the classical 
frequency $\omega(E)$ with small $h$ dependent corrections.  This should be contrasted with higher dimensions ${d>1}$, for which the level spacing ${\propto h^{d{-}1}}$ is highly non-classical. 
For ${d=1}$, the only region where $h$ determines the level spacing $\propto |\log(h)|^{-1}$ is in the vicinity of the separatrix as implied by Eq.~(\ref{e31}).

The Heisenberg time is defined as the inverse of the mean spacing of the participating levels. For a ${d=1}$ system and away from the separatrix, the Heisenberg time is merely the period of classical oscillations. Thus in the case of a Zero preparation we have ${\omega_{\tbox{osc}}  = \omega_J}$   
(the observed frequency is doubled due to Mirror symmetry).  Close to the separatrix, $\omega_{\tbox{osc}}$ becomes $h$ dependent as in Eqs.(\ref{e46}-\ref{e47}). This prediction is confirmed by the numerics (see \Fig{f7}), including the non-symmetry related factor~$2$ 
that distinguishes the Edge from the Pi preparation.  

Assume the system is prepared in some state $\psi$, e.g. a Gaussian-like SCS. We define $M$ as in Eq.~(\ref{e48}), implying that $\psi$ is roughly a superposition of $M$ energy states. If the energy levels are equally spaced the motion is strictly periodic.  Otherwise it is quasi-periodic. Our aim is to trace the non-classical behavior in the RMS of the temporal fluctuations of an observable $A$, say of the FringeVisibility as defined in Eq.~(\ref{e59}).      

Before proceeding, it should be made clear that the RMS of the fluctuations of any observable $A$ 
in a {\em classical} simulation (i.e. classical propagation of a single trajectory) is non-zero and characterized by its power spectrum $\tilde{C}_{\tbox{cl}}(\omega)$.  However, the RMS of the fluctuations in the {\em semiclassical} evolution (i.e. classical, leading-order propagation of a cloud of trajectories emulating the Wigner function) goes to zero due to the ergodic-like spreading of the wave-packet.  In contrast, the RMS of the fluctuations in the {\em quantum} evolution (corresponding to the full propagation to all orders, of the Wigner distribution) depends on~$M$. This dependence on~$M$ can be figured out by expanding the expectation value in the energy basis 
\beq
\bra A \ket_t = \sum_{\nu,\mu} \psi_{\nu}^*\psi_{\mu} A_{\nu\mu} \eexp{i(E_{\nu}-E_{\mu})t},
\eeq
where $\psi_{\nu} = \langle E_{\nu} |\psi\rangle$. 
The time average of this expectation value 
is $\overline{\bra A \ket_t} = \sum_\nu p_{\nu}A_{\nu,\nu}$ 
where ${p_{\nu} = \mbox{P}(E_{\nu})}$ of Eq.~(\ref{e42}).
This average has a well-defined $h$-independent classical limit. 
But if we first square, and then take the time average we get  
\be{7070}
\overline{\bra A \ket_t^2} = \sum_{\nu,\mu} p_{\nu} p_{\mu} |A_{\nu,\mu}|^2 ~.
\eeq
The matrix elements can be evaluated semiclassically using the well know 
relation $|A_{\nu,\mu}|^2 = \tilde{C}_{\tbox{cl}}(E_{\nu}{-}E_{\mu})/(2\pi\varrho)$, 
where $\varrho$ is the mean level spacing
(see Eq.(6) of Ref.\cite{lds} and references therein). 
For presentation purpose it is convenient to visualize $\tilde{C}_{\tbox{cl}}(\omega)$ 
as having a rectangular-like lineshape of width $\omega_{\tbox{cl}}$, 
such that its total area is ${\tilde{C}(0)\times \omega_c}$. 
It is also convenient to define the dimensionless bandwidth~$b$ 
as the spectral width of $\tilde{C}\tbox{cl}(\omega)$ divided 
by the mean level spacing, namely ${b=\varrho\omega_c}$.

Using the semiclassical estimate for the matrix elements, we consider the outcome of Eq.(\ref{e7070}),  in two limiting cases. If the energy spread of the wave-packet is smaller than the spectral 
bandwidth, we can factor out $\tilde{C}_{\tbox{cl}}(0)/(2\pi\varrho)$, 
and carry out the summation ${\sum p_{\nu} p_{\mu}=1}$, leading to  
\beq
\mbox{RMS}\left[\bra A \ket_t \right] 
= \left[\frac{1}{b} \int \tilde{C}\tbox{cl}(\omega) d\omega\right]^{1/2}.
\eeq
For integrable one-dimensional systems ${b\sim 1}$ reflects that only nearby levels are coupled.
A semiclassically large bandwidth ${b \propto \hbar^{1{-}d}}$ 
is typical for chaotic systems, which is not the case under consideration.  
Therefore we turn to the other possibility, in which 
the energy spread of the wave-packet is large compared 
with the spectral bandwidth. In such case we can make 
in Eq.(\ref{e7070}) the replacement ${p_{\nu}\mapsto 1/M}$, 
and consequently the sum ${\sum |A_{\nu,\mu}|^2}$ equals $M$ times 
the area of $\tilde{C}_{\tbox{cl}}(\omega)$, leading to 
\be{73}
\mbox{RMS}\left[\bra A \ket_t \right] 
= \left[\frac{1}{M} \int \tilde{C}\tbox{cl}(\omega) d\omega\right]^{1/2}.
\eeq
This is the same as the {\em classical} result but suppressed by factor $1/\sqrt{M}$. 
Note again that the {\em semiclassical} result is always zero, 
and corresponds formally to ${M=\infty}$.

Consider now the RMS of $S_x(t)$.  For TwinFock preparation it follows from Eq.(\ref{e4848}) 
that ${M\propto N}$ and hence Eq.(\ref{e73}) implies $1/N^{1/2}$ suppression of the RMS. 
For coherent preparations $\tilde{C}\tbox{cl}(\omega)$ becomes $N$~dependent too,   
and consequently from the discussion after Eq.~(\ref{e50})  
it follows that the RMS is a function of the semiclassical ratio Eq.~(\ref{e63}), 
and multiplied by $1/N^{1/4}$ suppression factor that {\em spoils} the semi-classical scaling. 
This is confirmed by the numerics (\Fig{f8}). 
If the dynamics is very close to the separatrix the classical 
fluctuations are ${\mathcal{O}(1)}$ and therefore the quantum result is  
${\mbox{RMS}\left[ S_x(t) \right] \approx 1/\sqrt{M}}$.  
The implication for the on-separatrix coherent preparations, Pi versus Edge, is striking: 
Substitution of Eqs.(\ref{e49}-\ref{e50}) into Eq.(\ref{e73}) leads to 
\be{74}
\mbox{RMS}\left[ S_x(t) \right] 
\ \ \sim \ \ \left\{\amatrix{ 
N^{-1/4}  &  \mbox{for Edge}  \cr
(\log(N))^{-1/2} &  \mbox{for Pi}
}\right. 
\eeq
Thus, convergence to classicality is far more rapid for the Edge-preparation 
than it is for the Pi-preparation, even though both lie on the separatrix. 
With the Pi preparation, even if $N$ is very large (small ``$h$"), 
quantum fluctuations still remain pronounced. 
In fact, from Eq.~(\ref{e49}) it follows that the fluctuations in the Pi case 
are mainly sensitive to the strength $u$ of the interaction.  

Finally, we mention that the analysis of fluctuations above 
is somewhat related to the discussion of thermalization in Ref.\cite{maxim},   
and we would like to further connect it with the observation of 
collapses and quantum revivals as discussed e.g. in Ref.\cite{scully}.
Relating to the LDOS, as defined in Section~V, we note that 
the collapse time is the semiclassical time which is determined 
by the the width of the classical envelope, while the revival time 
is related to the spacing between the spectral lines. 
The latter can be calculated using the formula 
\beq
t_{\tbox{revival}} \ \ = \ \ 2\pi \left[dE_{\nu}/d\nu\right]^{-1}
\eeq
with the WKB estimate for $E_{\nu}$ in Section~III, 
leading in the separtatrix region to $\sim 2\pi/\omega_{\tbox{x}}$.

\section{Conclusions}

To conclude, we have applied a semiclassical phase-space picture to the analysis of the one-particle coherence loss and buildup in the bosonic Josephson junction, described by the two-site BHH. The simplicity of the classical phase space of the dimer  allows for its semi-analytic WKB quantization. Thus, closed semiclassical results are obtained for the local density of states of the various initial preparations, providing useful insights for the associated quantum evolution.

Within the framework of mean-field theory (MFT), the dynamics is obtained by evolving a {\em single} point in phase space, using the Gross-Pitaevskii (GP) equation, which in this context is better known as the discrete nonlinear Schr\"odinger (DNLS) equation. 
By contrast, the truncated Wigner phase-space method evolves an {\em ensemble} of points according to the DNLS equation, and thereby takes into account the non-linear squeezing or stretching of the distribution. 

In the semiclassical treatment the quantum state in any moment is regarded  as a ``mixture" of wavefunctions $\psi_i$ rather 
than a single $\overline{\psi}_i$.  It is worth noting that the stationary solutions of the DNLS equation are simply the fixed points of the Hamiltonian flow. The small oscillations obtained by linearization around these fixed points are the so-called Bogoliubov excitations. The typical oscillation frequency of the Bloch vector  generally approaches the classical frequency as $N$ is increased keeping $u$ fixed. However, in the vicinity of the separatrix convergence to the (vanishing) classical frequency is logarithmically slow, as found via WKB quantization.

\begin{figure}[h!]

\centering
\includegraphics[clip,width=0.4\textwidth]{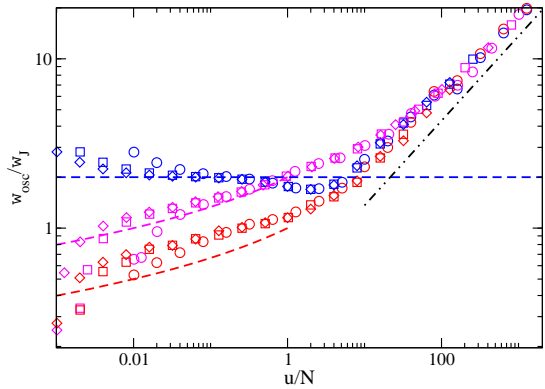}  

\caption{(Color online)
The mean frequency of the $S_x(t)$ oscillations versus $u/N$ for
for $N=100 \,(\circ),500\, (\Box),$ and $1000\,(\diamond)$ particles.
The preparations are (upper to to lower sets of data points):  
Zero (blue), Edge (magenta), and Pi (red).
The doubled Josephson frequency $2\omega_J$ 
is marked by a dashed blue line. 
The theoretical predictions of Eq.~(\ref{e46}), 
doubled due to mirror symmetry,  
are represented by red and magenta dashed lines, 
while Eq.~(\ref{e47}) for ${u/N\gg 1}$ is 
represented by black dash-double-dotted line 
(there is one fitting parameter as explained there).
}
\label{f7}
\end{figure}

Based on the ratio between the width of the semiclassical distribution for SCS and the width of the separatrix phase-space region, we find that the long-time FringeVisibility of an initially coherent state in the Josephson interaction regime, has a $u/N$ dependent value (\Fig{f7}). The functional dependence on $u/N$ varies according to the preparation. In particular, whereas a Zero relative-phase preparation remains roughly Gaussian (i.e. phase-locked) throughout its motion, thereby justifying the use of MFT for the description of Josephson oscillations around it, a Pi relative-phase SCS squeezes rapidly  and its relative-phase information is lost \cite{Vardi01,Boukobza09}. In contrast, starting from fully separated modes, the phase distribution in the Josephson regime assumes a non-uniform profile, peaked at $\varphi=0$, yielding a universal FringeVisibility value of $~1/3$ \cite{Boukobza09a}.

Focusing on two types of coherent preparations in the vicinity of the separatrix we find significant differences in their $M$ dependence on ${(u;N)}$.  The Pi SCS preparation (with vanishing population imbalance and a $\pi$ relative-phase) 
exhibits $u$~dependent fluctuations, whereas the Edge SCS (having a comparable 
population imbalance but located elsewhere along the separatrix) exhibits $N$~dependent fluctuations. Only in the latter case is the classical limit approached easily by taking large $N$ at fixed $u$.


\ \\

\noindent {\bf Acknowledgments:} 
We thank Issac Israel for preparing a convenient code 
for the classical simulations during his visit in BGU. 
DC and TK acknowledge support from the USA-Israel Binational Science Foundation (Grant No.2006021).
AV acknowledges support from the Israel Science Foundation (Grant 582/07),  
the USA-Israel Binational Science Foundation (Grant No.2008141), 
and the National Science Foundation through a grant for the Institute for Theoretical Atomic, Molecular, 
and Optical Physics at Harvard University and the Smithsonian Astrophysical observatory.
KSM, MH, and TK acknowledge support of the DFG within the Forschergruppe 760.


\appendix

\begin{figure}[h!]

\centering
\includegraphics[width=0.45\hsize]{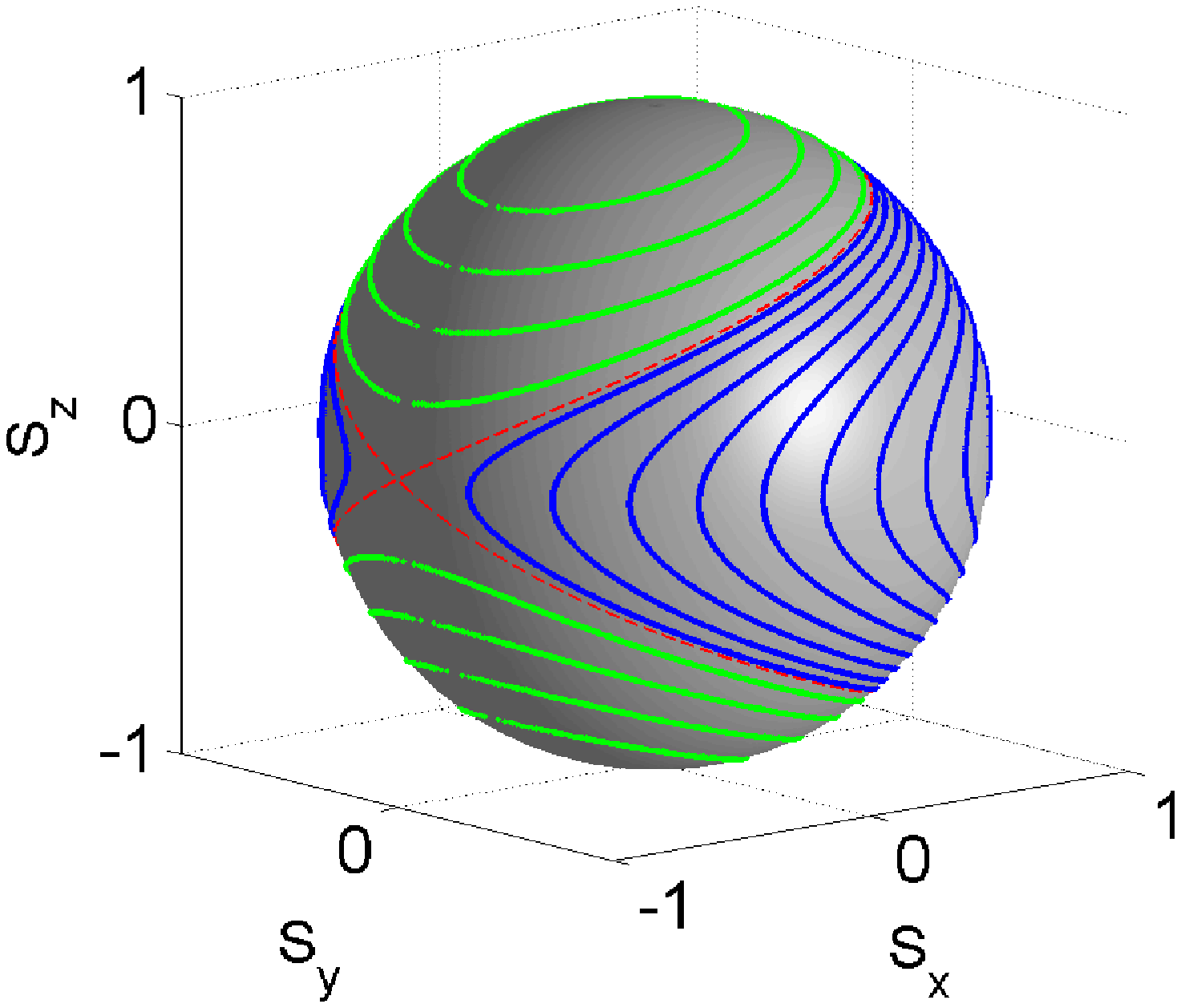} 
\includegraphics[width=0.45\hsize]{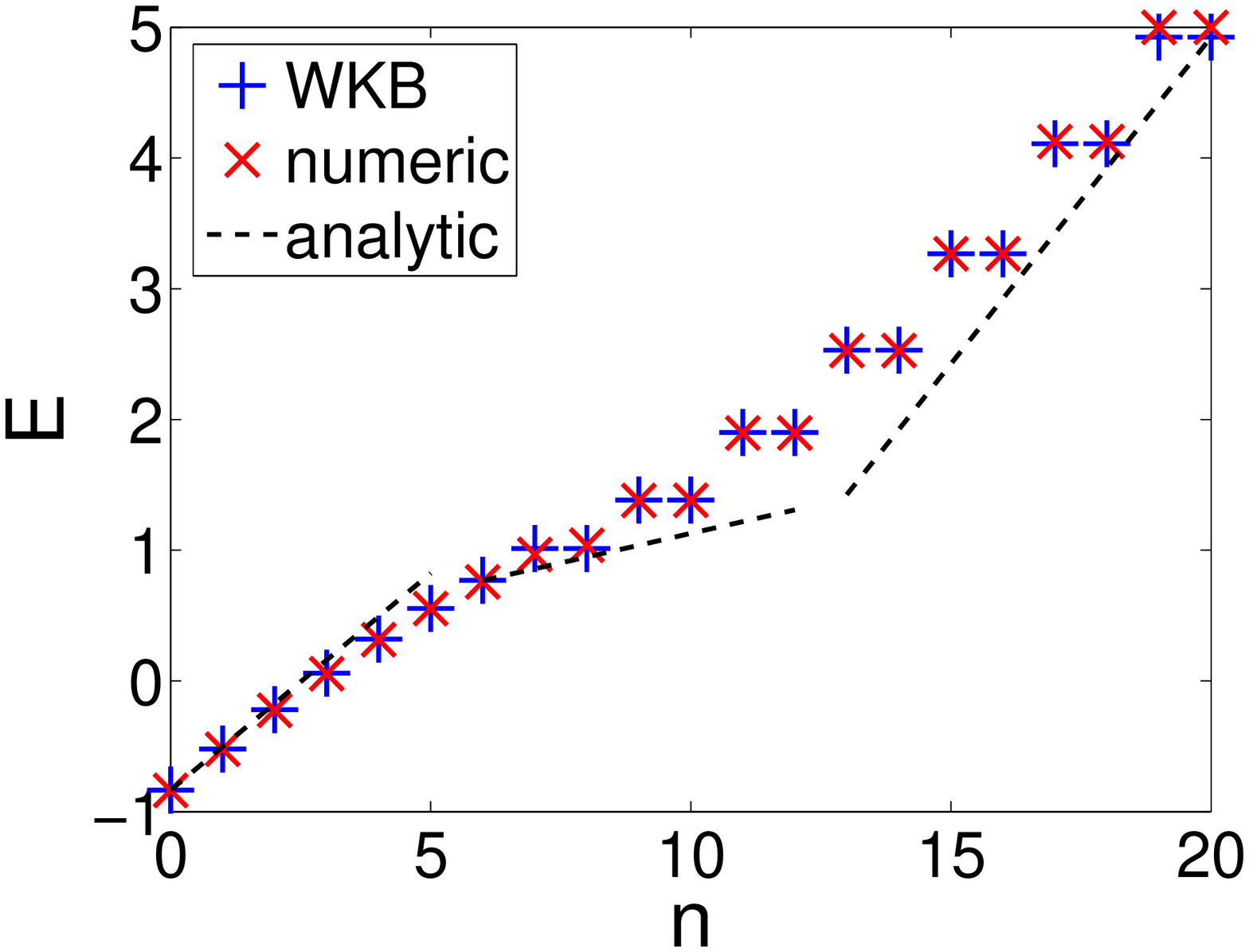}

\caption{
(Color online)  
Contour lines for $u>2$. Sea levels are colored blue, Island levels are colored green, 
and the Separatrix is colored red (left panel).
Energy spectrum for $N={20}$ and $u=10$. 
WKB energies (red x) are compared with exact eigenvalues (blue +).   
Dashed lines indicate slopes  $\omega_J$ for low energies, 
$\omega_x$ for near-separatrix energies, 
and $\omega_{+}$ for high energies (right panel).
}

\label{f1}
\end{figure}

\rmrkE{
\section{An $L$ site system with $N$ bosons}
\label{sA}

A bosonic $L$ site system has formally the same Hilbert space 
as that of $L$ coupled (harmonic) oscillators, with an additional constant 
of motion $N$ that counts the total number of quanta (rather than the total energy).
The creation operator $\hat{a}_i^{\dagger}$ corresponds to a raising operator and 
the occupation of the $i$th site $\nJ_i= \hat{a}_i^{\dagger}\hat{a}_i$ 
corresponds to the number of quanta stored in the $i$th mode. 
The one-particle states of an $L$ site system form 
an $L$ dimensional Hilbert space. The set of unitary 
transformations (i.e. generalized rotations) within this space is the $SU(L)$ group.
For $N$ particles in $L$ sites, the dimensionality of the Hilbert space is 
${\mbox{dim}(N)=(N{+}1)!/[(L{-}1)!(N{-}L{+}2)!]}$. For example for $L=2$ we have 
${\mbox{dim}(N)=N{+}1}$ basis states ${|n_1,n_2\rangle}$ with ${n_1+n_2=N}$. 
We can rotate the entire system using $\mbox{dim}(N)$ matrices. Thus we obtain a 
$\mbox{dim}(N)$ representation of the $SU(L)$ group. By definition these rotations 
can be expressed as a linear combination of the $SU(L)$ generators ${{J}}_{\mu}$, and they 
all commute with the conserved particle number operator~$\hat{N}$.
The many-body Hamiltonian $\mathcal{H}$ may contain ``nonlinear" 
terms such as ${{{J}}_{\mu}^2}$ that correspond to interactions 
between the particles. Accordingly, for an interacting 
system,  $\mathcal{H}$ is not merely a rotation. However, $\mathcal{H}$ still commutes with~$\hat{N}$, 
maintaining the fixed-number $\mbox{dim}(N)$ subspace.

In the semiclassical framework the dynamics 
in phase space is generated by the Hamilton equations 
of motion for the action-angle variables $\dot{\nJ}_i$ and $\dot{\varphiJ}_i$.
These are the ``polar coordinates" that describe each of 
the oscillators. It is common to define the complex coordinates   
\beq
\Psi_i \ \ \equiv \ \  \sqrt{\nJ_i} \, \eexp{i\varphiJ_i}
\eeq
(representing a single point in phase space).
This is the classical version of the destruction operator $\hat{a}_i$.
The equation for $\dot{\Psi}_i$ is  
the discrete nonlinear Schr\"odinger (DNLS) equation:
\beq
i\frac{d\Psi_i}{dt} \ \ = \ \ \Big(\epsilon_i + U|\Psi_i|^2 \Big) \Psi_i 
- \frac{K}{2}\left(\Psi_{i+1}+\Psi_{i-1} \right), 
\eeq
which is the space-discretized version of 
\beq
i\frac{d\Psi(x)}{dt} \ \ = \ \ \Big[V(x) + g_s|\Psi(x)|^2  -\frac{1}{2\mass}\nabla^2 \Big] \Psi(x)
\eeq
This looks like the  Gross-Pitaevskii (GP) equation, 
but strictly speaking it is {\em not} the GP~equation.
The GP~equation is the outcome of a mean-field theory (MFT):  
it is not an equation for $\Psi_i$, 
but an approximated equation for the mean-field $\overline{\Psi}_i$. 
We further clarify this point in the next paragraph. 

Within the framework of the semiclassical treatment the quantum state is described as a distribution 
of points in phase space. This approach goes beyond the conventional MFT approximation:  
MFT essentially assumes that the state of the system remains coherent throughout its evolution. 
Such a state corresponds to a Gaussian-like distribution is phase space (``minimal wave-packet") 
and accordingly it is fully characterized 
by  ``center of mass" coordinates ${(\overline{\varphi}_i,\overline{n}_i)}$
that are defined through the mean field  
\beq
\overline{\Psi}_i = \mbox{Mean Field}  \equiv \langle \Psi_i \rangle
\eeq
(representing the center of a wave-packet).
To the extent that the MFT assumption can be trusted, 
the equation of motion for $\overline{\Psi}_i$ is the DNLS / GP.
Indeed, if $u=0$ there is no nonlinear spreading 
and the MFT description becomes exact. 
The approximation holds well in the weak-interaction Rabi regime and for some 
regions of phase-space (e.g. around the ground state) in the strong-interaction 
Josephson regime, but in the nonlinear regions of phase-space MFT becomes 
too crude to provide an accurate description of the dynamics.

\section{The structure of phase space}
\label{sB}

The energy contours in a representative case are displayed in \Fig{f1}.
Considering a section along the big circle ${\varphi=0,\pi}$,  
it is convenient to take ${-\pi<\theta<+\pi}$ 
instead of ${0<\theta<\pi}$. Along this section the energy 
is ${E(\theta)=(NK/2)f(\theta)}$ where 
\beq
f(\theta) \ &\equiv& \ 
\frac{1}{2} u (\cos\theta)^2 - \varepsilon \cos\theta - \sin\theta\,
\\
f'(\theta) \ &=& \ 
-\frac{1}{2} u \sin(2\theta) + \varepsilon \sin\theta - \cos\theta,
\\ 
f''(\theta) \ &=& \ 
-u \cos(2\theta) + \varepsilon \cos\theta + \sin\theta\ .
\eeq
All extremal points of the Hamiltonian are located along this section
and are determined by the equation ${f'(\theta)=0}$.
The number of solutions depends on $u$ and on the bias $\varepsilon$.

For $u<1$, there are two fixed points: a minimum at ${\theta_{-}}$ 
at the bottom of the sea and a maximum at the opposite point  ${\theta_{+}}$. 
For $\varepsilon=0$ the two fixed points are: 
\beq
\theta_{-} =& \pi/2  \ \ \ \ \ \ \ &\mbox{[$\varphi{=}0$ point]},
\\
\theta_{+} =& -\pi/2 \ \ \ \ \ \ \ &\mbox{[$\varphi{=}\pi$ point]},
\eeq
with corresponding energies $E_{\pm}=\pm(N/2)K$.

For $u>1$, provided $\abs{\varepsilon}<\varepsilon_c$, 
the upper fixed point bifurcates into a saddle point $\theta_{\rm{x}}$ 
and two stable maxima $\theta_{1}$ and $\theta_{2}$. 
The value of the critical bias $\varepsilon_c$ is derived 
in the next paragraph. For $\varepsilon=0$ the four fixed points are: 
\beq
\theta_{-} =& +\pi/2,   \\
\theta_{\rm{x}} =& -\pi/2,   \\
\theta_{1,2} =& -\arcsin(1/u), 
\eeq
with corresponding energies $E_{-}$, $E_{\tbox{x}}$, and $E_{+}$ 
that are given by Eqs.(\ref{e12}-\ref{e14}). 
We can also determine the borders of the separatrix by solving 
the equation ${f(\theta_{1',2'}) = f(\theta_{\rm{x}})}$. 
Thus we get the following expressions for the outer borders of the islands 
\beq
\theta_{1',2'} \ \ = \ \ \arcsin(1-(2/u)) .
\eeq

The value of the critical bias Eq.~(\ref{e11}) is obtained 
by solving the equation ${f'(\theta)=0}$ together with ${f''(\theta)=0}$.
An equivalent set of questions is 
\beq
\sin(\theta)f''(\theta) - \cos(\theta)f'(\theta)  =   1+u\sin^3(\theta)  =  0,
\\
\cos(\theta)f''(\theta) + \sin(\theta)f'(\theta)  =   \varepsilon - u\cos^3\theta  = 0.
\eeq
leading to the solution
\beq
\theta_{\rm{x}} \ \ &=& \ \ -\arcsin((1/u)^{1/3}) 
\\
\varepsilon_c \ \ &=& \ \ \left(u^{2/3}-1\right)^{3/2}
\eeq
For $\varepsilon_c=\varepsilon$, one island has zero area while
the other has a critical area $A_c$. For completeness we 
derive an estimate for $A_c$, that we have used in \cite{SmithMannschott09}.
The island is defined by the equation $f(\theta,\varphi)=\const$., where
\beq
f(\theta,\varphi) \ \ = \ \  \frac{1}{2} u (\cos\theta)^2 - \varepsilon \cos\theta  - \sin\theta\cos\varphi.
\eeq
The critical island is defined by the equation ${f(\theta,\varphi)=f_c}$ 
where according to the previous paragraph 
\beq
f_c \ \ \equiv \ \ f(\theta_{\rm{x}};\varepsilon_c) \ \ = \ \ -\frac{1}{2}u +\frac{3}{2}u^{1/3}.
\eeq
The outer turning point $\theta_{1'}$ is identified 
as the second root of the equation $f(\theta) = f_c$. 
Defining $s \equiv \cos\theta$ the equation takes the form  
\be{-1}
\frac{u^2}{4} s^4 
- [\varepsilon_c u ] s^3
+ [1-f_c u {+} \varepsilon_c^2] s^2
+ [2 \varepsilon_c f_c] s
+ [f_c^2{-}1] =0.
\eeq
One root with double degeneracy is obviously 
\beq
s_{\rm{x}} \ \ = \ \ \cos(\theta_{\rm{x}}) \ \ = \ \ (1-u^{-2/3})^{1/2}.
\eeq
Then we can get the third root by solving a quadratic equation, leading to 
\beq
s_{1'} \ \ = \ \ 4s_{\rm{x}}^3-3s_{\rm{x}}.
\eeq
The area of the critical island can be found numerically, 
and a very good approximation is $s_x^3$, namely  
\beq
A_c \ \ \approx \ \ 4\pi \ (1-u^{-2/3})^{3/2}.
\eeq

\section{The Wigner function of a spin}
\label{sC}

The Stratonovich-Wigner-Weyl correspondence (SWWC) 
associates with any Hermitian operator of a spin$^{(j)}$,  
a real sphere$^{(2j)}$ function $A_{\tbox{W}}(\Omega)$. 
Both spaces have the same dimension~$(2j{+}1)^2$, and the association is one-to-one.  
This appendix provides the reasoning and the practical 
formulas following \cite{Agarwal81,wignerfunc}.
 
The inner product of two operators is defined as $\trc[\hat{A}^{\dag}\hat{B}]$.  
An orthonormal-like set of projector-like operators, 
that are knows and the Stratonovich-Weyl operators, 
can be defined, such that  
\beq
\int \frac{d\Omega}{h} \hat{P}^{\Omega}  \ \  &=& \ \ \hat{\bm{1}},
\\
\trc\left[\hat{P}^{\Omega} \hat{P}^{\Omega}\right]  \ \  &=& \ \ \delta_j(\Omega-\Omega')
\eeq
where ${\Omega=(\theta,\varphi)}$, and the ``delta" function is 
\beq 
\delta_j(\Omega-\Omega') = \sum_{\ell=0}^{2j}\sum_{m=-\ell}^{\ell} Y^{\ell m}(\Omega)Y^{\ell m *}(\Omega')
\eeq
Accordingly any operator $\hat{A}$ can be represented by the phase-space function  
\beq
A_{\tbox{W}}(\Omega) \ \ = \ \ \trc[ \hat{P}^{\Omega} \ \hat{\rho} ].
\eeq
From the above definition it follows that the inner product 
of two Hermitian operators is given by a phase-space integral 
over the product of the corresponding Wigner functions as in Eq.(\ref{e1234}).

The actual construction of the $\hat{P}^{\Omega}$ is not a simple task.
One procedure is to start with the non-orthonormal set of coherent state 
projectors ${|\Omega\rangle\langle\Omega|}$ and to perform an ``orthogonalization",  
very similar to that employed in condense matter physics for the 
purpose of defining a Wannier basis. The final result is conveniently expressed as   
\beq
\hat{P}^{\Omega} = \sqrt{\frac{4\pi}{2j+1}} \sum_{l=0}^{2j}\sum_{m=-l}^{l}   {Y}^{lm}(\Omega)  \ \hat{T}^{lm} ,
\eeq
which is an orthogonal transformation
over the non-Hermitian mulitpole operators $\hat{T}^{lm}$ defined as
\be{-1}
\hat{T}^{lm}
=
\sum_{m',m''} (-1)^{j-m'}\sqrt{2l+1}
\left( \begin{array}{ccc} 
 j &  l &  j \\  -m' & m &  m'' 
\end{array}\right)
|m'\rangle\langle m''|
\eeq
We use here the Wigner $3j$ symbols. 
Consequently we can represent any operator $\hat{A}$ 
either by ${A_{lm}=\trc[(\hat{T}^{lm})^{\dag}\hat{A}] }$ 
or by ${A_{\tbox{W}}(\Omega)=\trc[\hat{P}^{\Omega} \hat{A}]}$.
In particular it follows that $\rho=|\psi\rangle\langle\psi|$ 
is represented by 
\beq
\rho_{\tbox{W}}(\Omega) 
= 
\sum_{l,m}
C_{lm}
Y^{lm}(\Omega),
\eeq
where
\be{-1}
C_{lm}
= 
\sum_{m',m''} (-1)^{j{-}m'}
\sqrt{4\pi\frac{2l{+}1}{2j{+}1}}
\left( \begin{array}{ccc} 
 j &  l &  j \\  -m' & m &  m'' 
\end{array}\right)
\psi_{m'}^* \psi_{m''} .
\eeq
For the coherent state $\rho = |j,j\rangle\langle j,j|$
the only term in the sum is ${m'=m''=j}$, 
and therefore all the $C_{lm}$ are zero except $m=0$.
Consequently, the Wigner function is 
\be{-1}
\rho_{\tbox{W}}(\Omega) 
=
\sum_{l=0}^{2j} 
(2j)! \sqrt{\frac{4\pi\,(2l+1)/(2j+1)}{(2j+l+1)!(2j-l)!}}
\ {Y}^{l,0}(\Omega).
\eeq
This function resembles a minimal Gaussian as 
one can see from the example in \Fig{f2}. 

}

\ \\ 
\begin{widetext}

\section{LDOS semiclassical calculation}
\label{sD}

Using the definition Eq.~(\ref{e42}) we 
can calculate the LDOS in the semiclassical 
approximation using Eq.~(\ref{e40}).  
The integral that should be calculated
in the case of Zero ($\varphi_{-}=0$)
and Pi ($\varphi_{\rm x}=\pi$) preparations is
\beq
\mbox{P}(E)
\ \ = \ \ 
\omega(E)
\int\int \frac{d\varphi dn}{2\pi} 
\ \delta\left(U {n}^2 - \frac{NK}{2} \cos(\varphi)-E\right)
\ \frac{1}{ab}\exp\left[ -\frac{(\varphi-\varphi_{-,{\rm x}})^2}{2 a^2} - \frac{n^2}{2b^2}\right],
\eeq
with $E=E_{\nu}$. Most of the contribution comes from 
the vicinity of the fixed point,  
so we can use a quadratic approximation:  
\beq
\mbox{P}(E)
\ \ = \ \ 
\omega(E)
\int\int \frac{d\varphi dn}{2\pi} 
\ \delta\left(U {n}^2 \pm \frac{1}{4}NK \varphi^2 - (E{-}E_{-,{\rm x}})\right)
\ \frac{1}{ab}\exp\left[ -\frac{\varphi^2}{2 a^2} - \frac{n^2}{2b^2}\right],
\eeq
Below we derive the following results for the Zero and Pi preparations: 
\beq
\mbox{P}(E)  &=&
2\exp\left[ -
\left(  \frac{1}{NU}+\frac{4}{K} \right)
(E-E_{-})\right]
\ \mbox{\bf I}_0\left[
\left( \frac{4}{K} - \frac{1}{NU} \right)
(E-E_{-})\right] ,
\\
\mbox{P}(E)  &=&
\frac{1}{\pi}\left(\frac{\omega(E_x)}{\omega_J}\right)
\ \exp\left[ 
\left( \frac{4}{K} -\frac{1}{NU} \right)
(E-E_{\tbox{x}})\right]
\ \mbox{\bf K}_0\left[
\left(\frac{4}{K}+ \frac{1}{NU}  \right)
|E-E_{\tbox{x}}|\right] .
\eeq
In order to simplify notations we define bellow $\epsilon$ 
as the difference $E{-}E_{-}$ or $E{-}E_{\tbox{x}}$.

In order to estimate the integral in the Zero case
we change to polar coordinates $r$ and $-\pi<t<\pi$: 
\beq
\varphi &=& [4\epsilon/NK]^{1/2} \ r \cos(t), \\
n &=& [\epsilon/U]^{1/2} \ r \sin(t),
\eeq
leading to 
\beq
\mbox{P}(E) 
\ \ &=& \ \ 
\omega_0
\int\int
\frac{rdrdt}{2\pi} 
\ \frac{2}{\omega_0}\delta\left(r^2-1\right)
\ \frac{1}{ab}\exp\left[ 
- (4\epsilon/NK)\frac{(\cos(t))^2}{2 a^2}r^2 
- (\epsilon/U)\frac{(\sin(t))^2}{2b^2}r^2
\right]
\\
\ \ &=& \ \
\frac{1}{2\pi} 
\int_{-\pi}^{+\pi}
dt
\ \frac{1}{ab}\exp\left[ 
- (4\epsilon/NK)\frac{(\cos(t))^2}{2 a^2} 
- (\epsilon/U)\frac{(\sin(t))^2}{2b^2}
\right]
\\
\ \ &=& \ \
\frac{1}{\pi} 
\int_{-\pi}^{+\pi}
dt
\ \exp\left[ 
- \left(\frac{2\epsilon}{K}+\frac{\epsilon}{2NU}\right)     
- \left(\frac{2\epsilon}{K}-\frac{\epsilon}{2NU}\right) \cos(2t)   
\right]
\\
\ \ &=& \ \
\exp\left[- \left(\frac{2}{K}+\frac{1}{2NU}\right)\epsilon \right]
\ 2\mbox{\bf I}_0\left[\left(\frac{2}{K}-\frac{1}{2NU}\right)\epsilon \right]
\eeq
It is important to realize that the semiclassical evaluation 
in the Zero case is valid only if the contour lines of the Gaussian 
intersect the contour lines of the eigenstates transversely.
Else, the Airy structure of the eigenstates should be taken into account, 
or perturbation theory rather than semiclassics should be used.
The asymptotic behavior of the Bessel function 
is ${\mbox{\bf I}_0(x) \approx \exp(x)/\sqrt{2\pi x}}$
and hence ${\mbox{P}(E) \sim \exp[-\epsilon/(NU)]}$. 
One observes that the tails of the LDOS
reflect the relatively slow decay 
of the Gaussian tails in the $n$ direction. 

In order to estimate the integral in the Pi case
we change to the coordinates $r$ and $-\infty<t<\infty$:
\beq
\varphi =& [4\epsilon/NK]^{1/2} \ r \sinh(t) \ \ \ \ \ \ &\mbox{or $\cosh(t)$ for $\epsilon<0$},     \\
n =& [\epsilon/U]^{1/2} \ r \cosh(t) \ \ \ \ \ \ &\mbox{or $\sinh(t)$ for $\epsilon<0$} ,
\eeq
leading to 
\beq
\mbox{P}(E)
\ \ &=& \ \ 
\omega(E)
\int\int
\frac{rdrdt}{2\pi} 
\ \frac{2}{\omega_0}\delta\left(r^2-1\right)
\ \frac{1}{ab}\exp\left[
- (4|\epsilon|/NK)\frac{(\sinh(t))^2}{2 a^2}r^2 
- (|\epsilon|/U)\frac{(\cosh(t))^2}{2b^2}r^2 
\right]
\\
\ \ &=& \ \
\frac{1}{2\pi}\left(\frac{\omega(E)}{\omega_0}\right)
\int_{|t|<\frac{1}{\pi}\left(\frac{\omega(E)}{\omega_0}\right)}
dt
\ \frac{1}{ab}\exp\left[             
- (4|\epsilon|/NK)\frac{(\sinh(t))^2}{2 a^2} 
- (|\epsilon|/U)\frac{(\cosh(t))^2}{2b^2}
\right]
\\
\ \ &=& \ \
\frac{1}{\pi}\left(\frac{\omega(E)}{\omega_0}\right)
\int_{|t|<\frac{1}{\pi}\left(\frac{\omega(E)}{\omega_0}\right)}
dt
\ \exp\left[ 
\pm \left(\frac{2|\epsilon|}{K}-\frac{|\epsilon|}{2NU}\right)     
- \left(\frac{2|\epsilon|}{K}+\frac{|\epsilon|}{2NU}\right) \cosh(2t)  
\right]
\\
\ \ &=& \ \
\frac{1}{\pi}\left(\frac{\omega(E)}{\omega_0}\right)
\exp\left[\pm \left(\frac{2}{K}-\frac{1}{2NU}\right)|\epsilon|  \right]
\ \mbox{\bf K}_0\left[\left(\frac{2}{K}+\frac{1}{2NU}\right)|\epsilon| \right].
\eeq
Note that without the Gaussian the result of the integral should be 
unity reflecting the proper normalization of the microcanonical state.
The upper cutoff of the $t$ integration reflects the finite size of phase space.
The cutoff prevents the singularity in the limit $\epsilon\rightarrow 0$. 
The Bessel function expression is obviously valid only outside of the 
cutoff-affected peak.
The asymptotic behavior of the Bessel function 
is ${\mbox{\bf K}_0(x) \approx \exp(-x)/\sqrt{2 \pi x}}$,  
and hence we get ${\mbox{P}(E)\sim \exp[-|\epsilon|/(NU)]}$ for ${\epsilon>0}$
and ${\mbox{P}(E)\sim \exp[-|\epsilon|/(K/4)]}$ for ${\epsilon<0}$.
One observes that the tails of the LDOS
for ${\epsilon<0}$ reflect the relatively rapid decay 
of the Gaussian tails in the $\varphi$ direction. 

Finally we point out that the calculation of the LDOS for 
the Edge and for the TwinFock preparations is much easier.
In the former case the delta function of ${\rho^{(\nu)}(n,\varphi)}$  
merely replaces the $n$~coordinate in ${\rho^{(\psi)}(n,\varphi)}$ 
by a constant proportional to ${E{-}E_{\tbox{x}}}$, 
leading to a Gaussian for $\mbox{P}(E)$.  
Similarly for the TwinFock preparation:  
\beq
\mbox{P}(E)
\ \ = \ \ 
\omega(E)
\int\int \frac{d\varphi dn}{2\pi} 
\ \delta\left(U {n}^2 - \frac{NK}{2} \cos(\varphi)-E\right) \ \delta(n) 
\ \ \propto  \ \ \left. \frac{\omega(E)}{(NK/2)|\sin(\varphi)|} \right]_{E}
\eeq
The later expression should be calculated for $(NK/2)\cos(\varphi)=E$ 
leading to Eq.(\ref{e4848}) in the main text. \\ 

\end{widetext}

\begin{figure}[h!]

(a) \hspace{0.3\hsize} (b) \hspace{0.3\hsize} (c) \hspace{0.3\hsize} \\
\centering
\includegraphics[width=0.32\hsize]{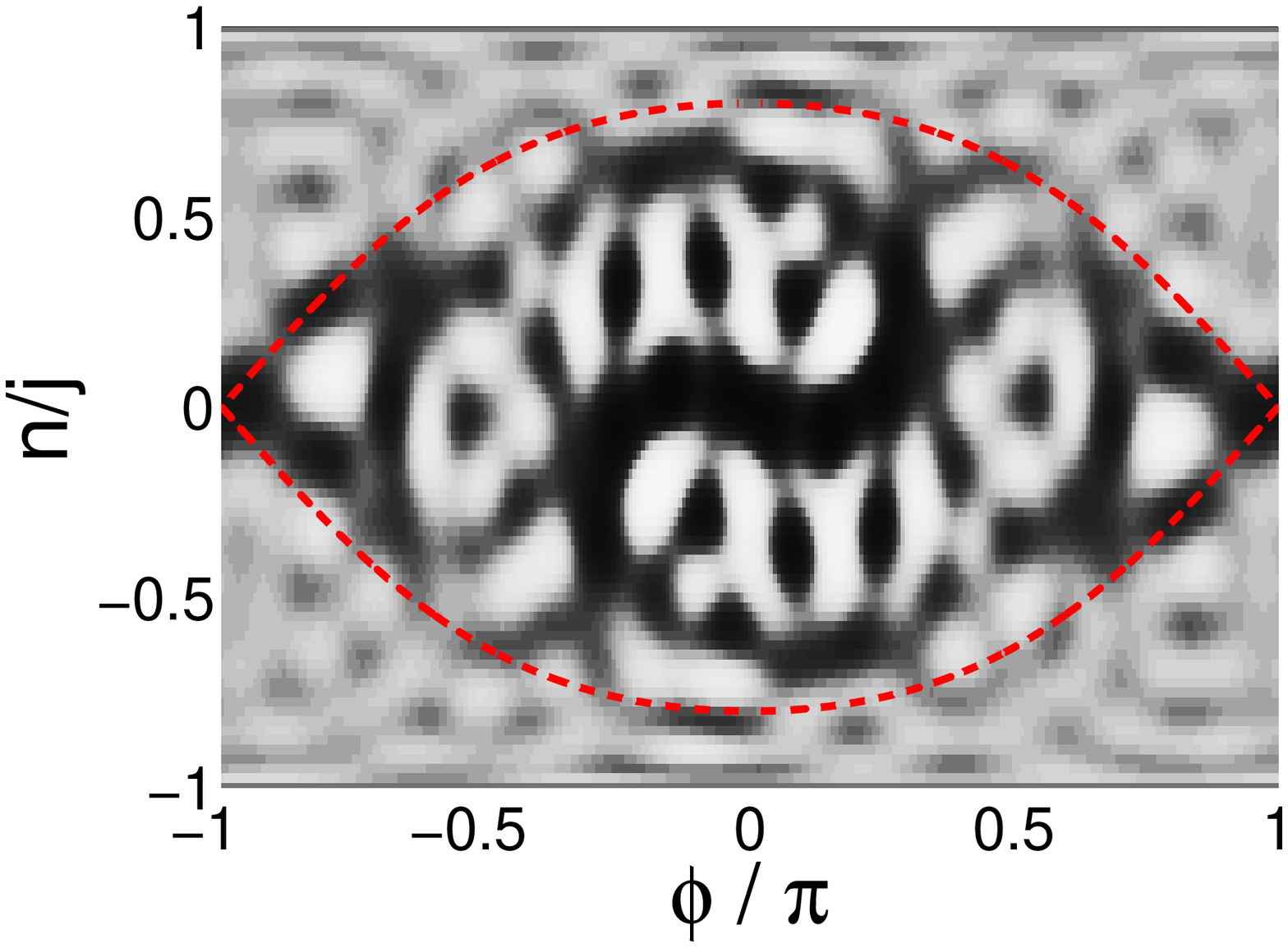}  
\includegraphics[width=0.32\hsize]{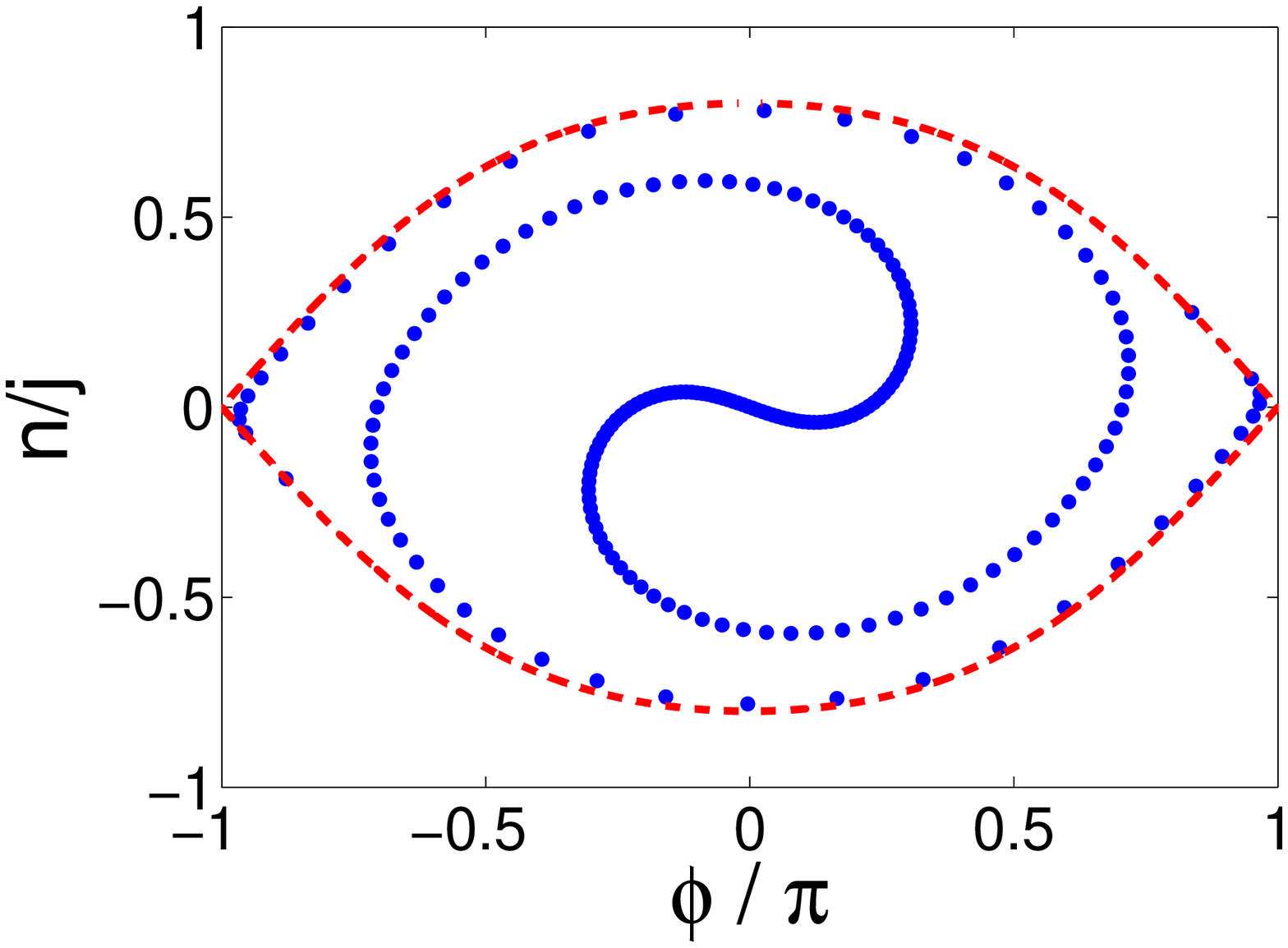} 
\includegraphics[width=0.32\hsize]{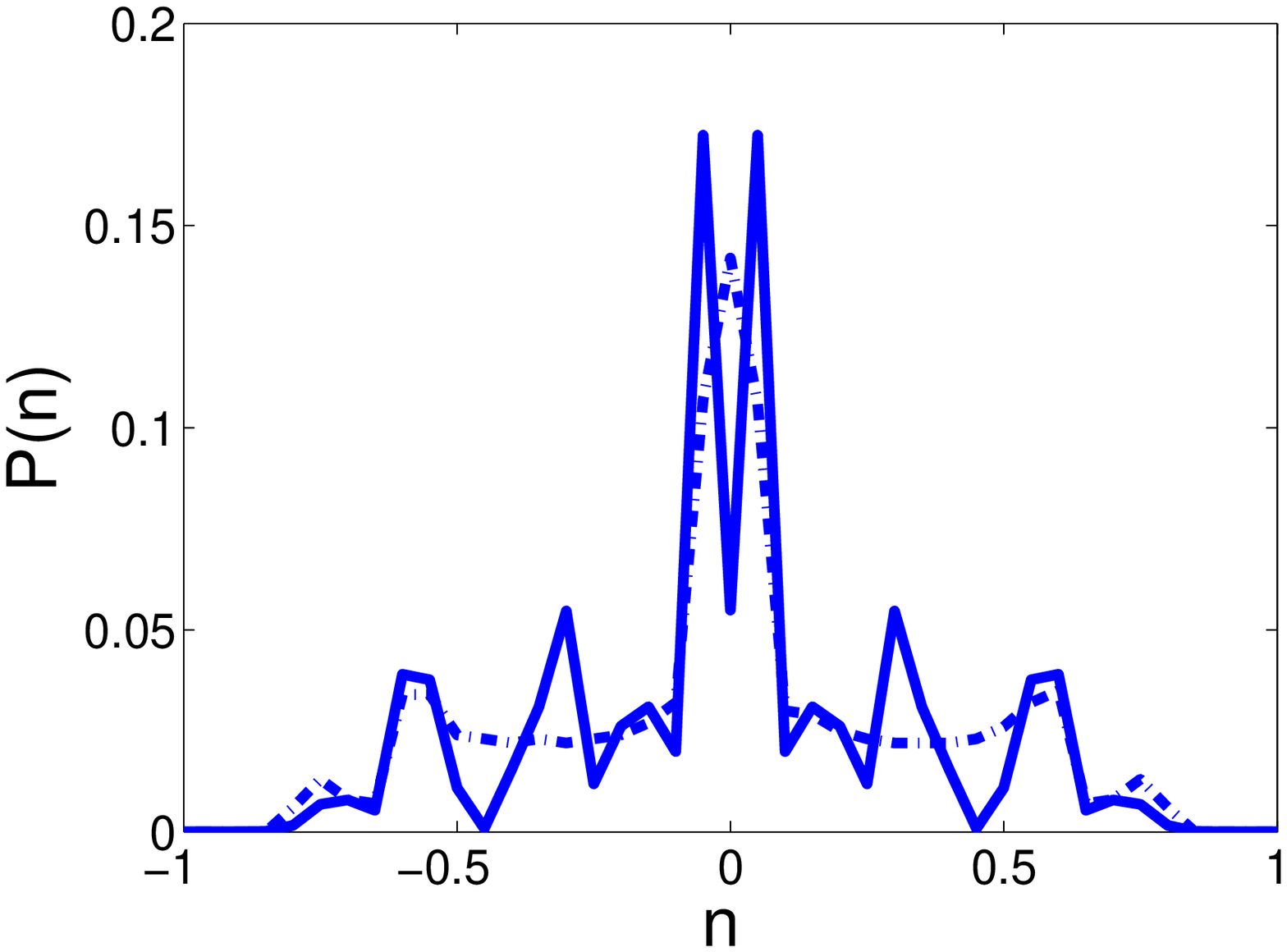} 

\caption{(Color online)
The evolving quantum state of $N=40$ bosons with $u=5$ for TwinFock ($n=0$) preparation.
Here and below the units are such that ${K=1}$. The time is $t=4$. 
(a) Wigner function of the evolved quantum state.
(b) Semiclassical evolved state.
(c) Occupation statistics, with the semiclassical result shown as dashed-dotted line. 
}
\label{f3}
\end{figure}

\rmrkE{
\section{Rabi-Josephson oscillations, MFT and semiclassical perspectives}

{\bf MFT.--} 
For $u=0$, the Hamiltonian (\ref{e5}) merely generates rotations in phase space. 
Consequently, coherent states remain Gaussian-like throughout their motion, 
resulting in Rabi oscillations of the population between the sites.
This is the case where MFT gives exact results. 
For $u\ne0$,  MFT maintains the assumption that the system remains in a coherent state at any instant during its evolution and that propagation only serves to displace this Gaussian distribution on the Bloch sphere.  Thus, MFT corresponds to the classical dynamics of a ``point" in phase space. In particular, this implies that the occupation statistics is binomial at any time.

Nonlinear effects cannot be neglected once the wave-packet is stretched along the phase-space energy contour lines. Then MFT no longer applies, but the semiclassical approximation still works well for the description of such squeezing. For example, by projecting the evolving semiclassical distribution onto~$n$ we obtain the number distribution $P_t(n)$ \cite{SmithMannschott09} whose line shape reflects the phase space structure, with caustics at the borders of forbidden regions (see \Fig{f3}).

{\bf Semiclassics. --}
For $u=0$, all trajectories  have the same topology and frequency and a Gaussian-like wave-packet that is launched (say) at the NorthPole executes Rabi oscillation between the two wells. If~$u$ is non-zero but small, then all the trajectories still have the same topology, but $\omega(E_{\nu})$ is $\nu$ dependent. The anharmonic behavior is due to the spectral stretch $\Delta\omega_{\tbox{osc}}$.  
Accordingly, we distinguish between the harmonic stage (${ t < 1/ \Delta\omega_{\tbox{osc}}}$) 
and the anharmonic stage (${t > 1/\Delta\omega_{\tbox{osc}} }$) in the time evolution.

For $u>1$, which we call ``the Josephson regime", a separatrix emerges and accordingly there are two types of stable coherent oscillations, depending on the initial population imbalance \cite{Smerzi97,Albiez05}: Small oscillations around the ground state, lying in the bottom of the sea; and self-trapped oscillations around the top of either island. 

For ${u > N^2}$ which we call ``the Fock regime", the area of the sea becomes less than a Planck cell, and therefore effectively disappears.  Our main interest in this work is the unstable motion along the separatrix for ${1\ll u \ll N^2}$ \cite{Vardi01,Khodorkovsky08,Boukobza09}. Such motion emerges for the Pi and for the Edge preparations of \Fig{f2}.

{\bf Beyond. --}
Quantum effects that are ignored by the semiclassical approximation are anharmonic beats, long-time recurrences, and the possibility of tunneling between the two islands. Associated with the recurrences are the long-time fluctuations, both in the occupation and in the FringeVisibility, as discussed in Section~VIII. 

}




\onecolumngrid

\clearpage

\begin{figure}[h!]

\centering
\includegraphics[clip,width=0.45\textwidth]{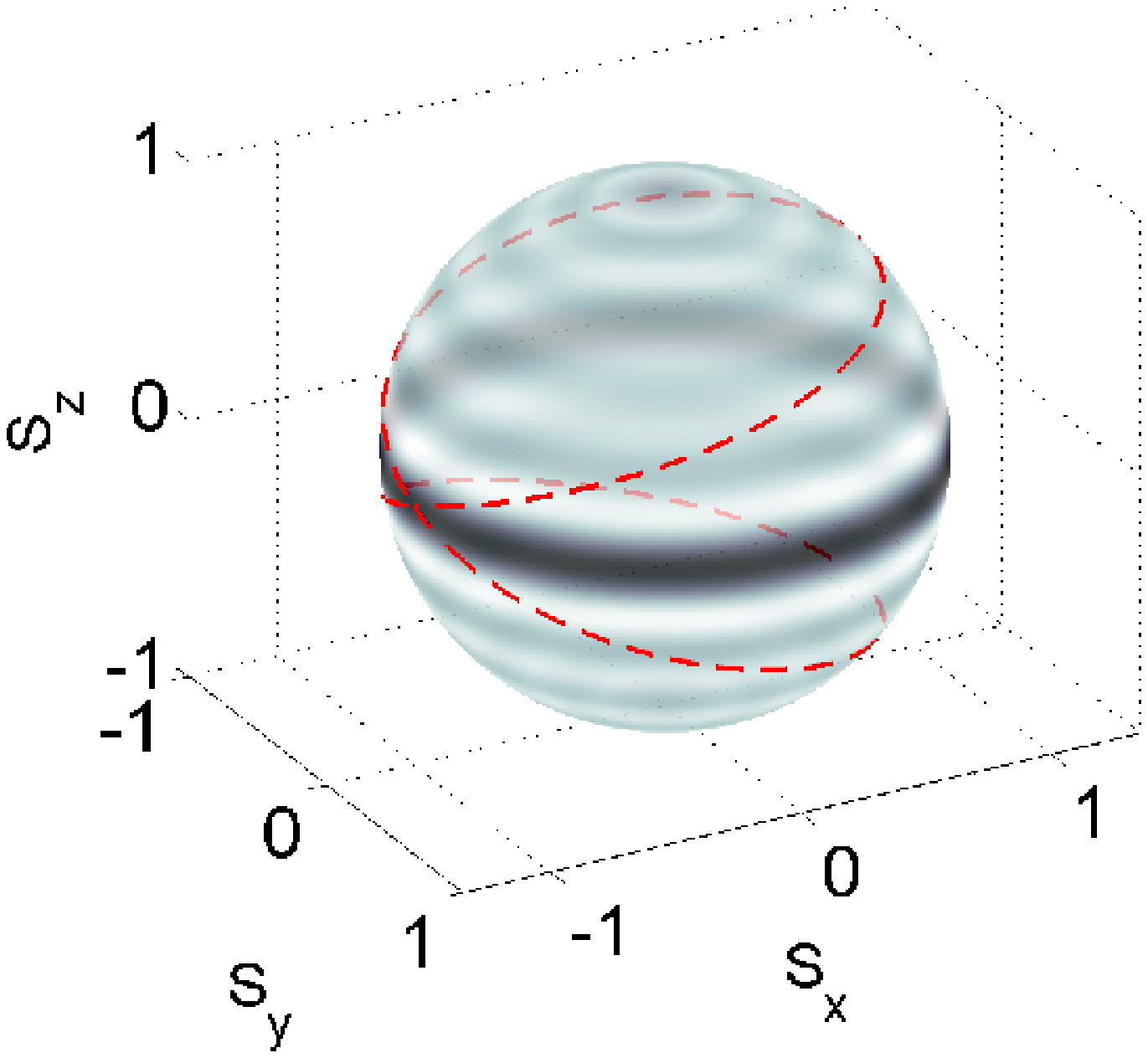} 
\includegraphics[clip,width=0.45\textwidth]{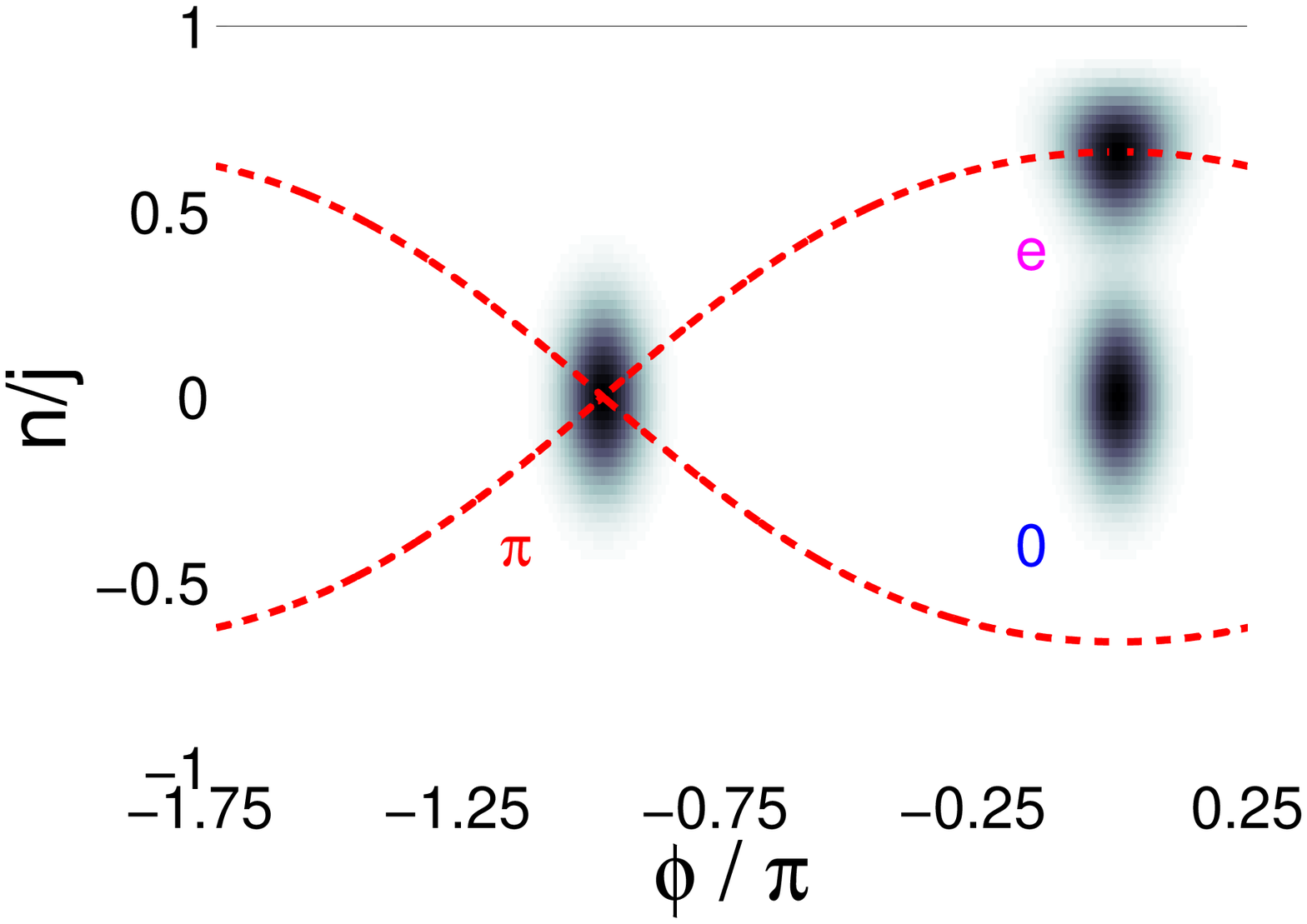} 

\caption{
(Color online) 
An illustration of TwinFock (${n=0}$) preparation (left), 
and of Pi ("$\pi$"),  Zero ("$0$") and Edge ("e") preparations (right) 
using Wigner plots on a sphere. The left panel is a 3D plot, 
while the right panel is a Mercator projection of the sphere 
using ${(\varphiJ,\nJ)}$ coordinates.
}
\label{f2}
\end{figure}

\clearpage


\clearpage

\begin{thebibliography}{99}


\bibitem[a]{nA}
The arXiv:1001.2120 version of this paper includes 3 extra pedagogically oriented 
appendices: Bosons in $N$~site system and MFT; 
Detailed analysis of phase space geometry in the $N{=}2$ case; 
The definition of a Wigner function of a spin; 
Concise discussion of the Rabi-Josephson oscillations in the 
semiclassical perspective.
   

\bibitem{Jaksch98}
D. Jaksch C. Bruder, J. I. Cirac, C. W. Gardiner, and P. Zoller,  Phys. Rev. Lett. {\bf 81}, 3108 (1998).

\bibitem{Greiner02a}
M. Greiner, O. Mandel, T. Esslinger, T. W. H{\"a}nsch, and I. Bloch,
Nature (London) {\bf 415}, 39 (2002).

\bibitem{BJM}
Y. Makhlin, G. Sch\"on, and A. Shnirman, Rev. Mod. Phys. {\bf 73}, 357 (2001); 
R. Gati and M. K. Oberthaler, J. Phys. B {\bf 40}, R61 (2007).

\bibitem{Leggett01}
Gh-S. Paraoanu {\it et al.}, J. Phys. B: At. Mol. Opt. Phys. {\bf 34}, 4689 (2001); 
A. J. Leggett, Rev. Mod. Phys. {\bf 73}, 307 (2001).

\bibitem{Javanainen86}
J. Javanainen, Phys. Rev. Lett. {\bf 57},  3164 (1986).

\bibitem{Dalfovo96}
F. Dalfovo, L. Pitaevskii, and S. Stringari, Phys. Rev. A {\bf 54}, 4213 (1996).

\bibitem{Zapata98}
I. Zapata, F. Sols, and A. J. Leggett, Phys. Rev. A {\bf 57}, 1050 (1998).

\bibitem{Smerzi97}
A. Smerzi, S. Fantoni, S. Giovanazzi, and R. S. Shenoy, Phys. Rev. Lett. {\bf 79}, 4950 (1997).

\bibitem{Cataliotti01}
F. S. Cataliotti, S. Burger, C. Fort, P. Maddaloni, F. Minardi,  A. Trombettoni, A. Smerzi, and M. Inguscio, Science {\bf 293}, 843 (2001).

\bibitem{Albiez05}
M. Albiez, R. Gati, J. F\"olling, S. Hunsmann, M. Cristiani, and M. K. Oberthaler, 
Phys. Rev. Lett. {\bf 95}, 010402 (2005).

\bibitem{Giovanazzi00}
S. Giovanazzi, A. Smerzi, and S. Fantoni, Phys. Rev. Lett.  {\bf 84}, 4521 (2000).

\bibitem{Levy07}
S. Levy, E. Lahoud, I. Shomroni, and J. Steinhauer, 
Nature {\bf 449}, 579 (2007); C. A. Sackett, Nature {\bf 449}, 546 (2007). 

\bibitem{Leggett98}
A. J. Leggett and F. Sols, Found. Phys.  {\bf 21}, 353 (1998).

\bibitem{Wright96}
E. M. Wright, D. F. Walls and J. C. Garrison Phys. Rev. Lett. {\bf 77}, 2158 (1996).

\bibitem{Javanainen97}
J. Javanainen and M. Wilkens, Phys. Rev. Lett. {\bf 78}, 4675 (1997); Phys. Rev. Lett. {\bf 81}, 1345 (1998).

\bibitem{Greiner02}
M. Greiner, O. Mandel, T. W. H\"ansch, and I. Bloch, Nature {\bf 419}, 51 (2002).

\bibitem{Jo07}
G.-B. Jo {\it et al.}, Phys. Rev. Lett. {\bf 98}, 030407 (2007).

\bibitem{Schumm05}
T. Schumm {\it et al.}, Nat. Phys. {\bf 1}, 57 (2005). 

\bibitem{Hofferberth07}
S. Hofferberth {\it et al.}, Nature (London) {\bf 449}, 324 (2007).

\bibitem{Widera08}
A. Widera, S. Trotzky, P. Cheinet, S. F\"olling, F. Gerbier, I. Bloch, V. Gritsev, M. D. Lukin, and E. Demler,  Phys. Rev. Lett. {\bf 100}, 140401 (2008).

\bibitem{Vardi01}
A. Vardi and J. R. Anglin, Phys. Rev. Lett. {\bf 86}, 568 (2001); 
J. R. Anglin and A. Vardi, Phys. Rev. A {\bf 64}, 013605 (2001).

\bibitem{Khodorkovsky08} 
Y. Khodorkovsky, G. Kurizki, and A. Vardi, Phys. Rev. Lett. {\bf 100}, 220403 (2008); Phys. Rev. A {\bf 80}, 023609 (2009).

\bibitem{Boukobza09}
E. Boukobza, M. Chuchem, D. Cohen, and A. Vardi, Phys. Rev. Lett. {\bf 102}, 180403 (2009).

\bibitem{SmithMannschott09}
K. Smith-Mannschott, M. Chuchem, M. Hiller, T. Kottos, and D. Cohen,
Phys. Rev. Lett. {\bf 102}, 230401 (2009).

\bibitem{GardinerZoller}
C. W. Gardiner and P. Zoller, {\it Quantum Noise}, 2nd ed. (Springer, New York, 2000).

\bibitem{Steel98}
M. J. Steel, M. K. Olsen, L. I. Plimak, P. D. Drummond, S. M. Tan, M. J. Collett, D. F. Walls, and R. Graham, Phys. Rev. A {\bf 58}, 4824 (1998).

\bibitem{Sinatra01}
A. Sinatra, C. Lobo, and Y. Castin, 
Phys. Rev. Lett. {\bf 87}, 210404 (2001); 
J. Phys. B: At. Mol. and Opt. Phys. {\bf 35}, 3599 (2002).

\bibitem{Carusotto01}
I. Carusotto, Y. Castin, and J. Dalibard, 
Phys. Rev. A {\bf 63}, 023606 (2001).

\bibitem{Hoffmann08}
S. E. Hoffmann, J. F. Corney, and P. D. Drummond,
Phys. Rev. A {\bf 78}, 013622 (2008).

\bibitem{Polkovnikov03}
A. Polkovikov,
Phys. Rev. A {\bf 68}, 053604 (2003).

\bibitem{Plimak03}
L. I. Plimak, M. Fleischhauer, M. K. Olsen, and M. J. Collett,
Phys. Rev. A {\bf 67}, 013812 (2003).

\bibitem{Deuar06}
P. Deuar and P. D. Drummond,
J. Phys. A: Math and Gen. {\bf 39}, 1163 (2006);
J. Phys. A: Math and Gen. {\bf 39}, 2723 (2006);

\bibitem{Isella06}
L. Isella and J. Ruostekoski,
Phys. Rev. A {\bf 74}, 063625 (2006).

\bibitem{HKG06}
M. Hiller, T. Kottos, and T.Geisel , Phys. Rev. A {\bf 73}, 061604(R) (2006); 
Phys. Rev. A {\bf 79}, 023621 (2009).


\bibitem{Midgley09}
S. L. W. Midggley, S. W\"uster, M. K. Olsen, M. J. Davis, and K. V. Kheruntsyan, 
Phys. Rev. A {\bf 79}, 053632 (2009).

\bibitem{Deuar09}
P. Deuar, 
Phys. Rev. Lett {\bf 103}, 130402 (2009).

\bibitem{Trimborn08}
F. Trimborn, D. Witthaut, and H. J. Korsch,
Phys. Rev. A {\bf 77}, 043631 (2008);
Phys. Rev. A {\bf 79}, 013608 (2009).

\bibitem{Mahmud}
K.W. Mahmud, H. Perry, and W.P. Reinhardt,
Phys. Rev. A 71, 023615 (2005).



\bibitem{Franzosi00}
Franzosi et al, Int. J. Mod.  Phys. B 14, 943 (2000).


\bibitem{Boukobza09a}
E. Boukobza, D. Cohen, and A. Vardi, Phys. Rev. A {\bf 80}, 053619 (2009).

\bibitem{Graefe07}
E.M. Graefe, H.J. Korsch, Phys. Rev. A 76, 032116 (2007); D. Witthaut, E. M. Graefe, and H. J. Korsch, Phys. Rev. A 73, 063609 (2006).

\bibitem{Agarwal81}
G. S. Agarwal, Phys. Rev. A. {\bf 24}, 2889  (1981).
J. P. Dowling, G. S. Agarwal,W. P Schleich, Phys. Rev. A {\bf 49}, 4101 (1994). 


\bibitem{wignerfunc} 
J.C. Varilly and J.M. Gracia-Bondia, Annals of Physics {\bf 190}, 107 (1989).
C. Brif and A. Mann, J. Phys. A {\bf 31}, L9 (1998).

\bibitem{Polkovnikov02}
A. Polkovnikov, S. Sachdev, and S.M. Girvin, 
Phys. Rev. A {\bf 66}, 053607 (2002).

\bibitem{Altman02}
E. Altman and A. Auerbach,
Phys. Rev. Lett. {\bf 89}, 250404 (2002).

\bibitem{Tuchman06}
A. K. Tuchman, C. Orzel, A. Polkovnikov, and M. Kasevich,
Phys. Rev. A {\bf 74}, 051601 (2006).

\bibitem{WN00}
B. Wu and Q. Niu, Phys. Rev. A {\bf 61}, 023402 (2000).

\bibitem{ELS85}
J. C. Eilbeck, P. S. Lomdahl, and A. C. Scott,
Physica D {\bf 16}, 318 (1985).

\bibitem{Stratonovich56}
R. L. Stratonovich,  Sov. Phys. JETP {\bf 31}, 1012 (1956).

\bibitem{lds}
D. Cohen and T. Kottos, 
Phys. Rev. E {\bf 63}, 36203 (2001). 

\bibitem{Boukobza10}
E. Boukobza, M. G. Moore, D. Cohen, and A. Vardi,
Phys. Rev. Lett. {\bf 104}, 240402 (2010).

\bibitem{maxim}
M. Rigol, V. Dunjko, and M. Olshanii, 
Nature {\bf 452}, 854 (2008).    

\bibitem{scully}
{\em Quantum Optics} by M.O. Scully and M.S. Zubairy,  
(Cambridge University Press, 1997), p.201.


\end{thebibliography}
\end{document}